\documentclass[]{aa}  
\usepackage{siunitx}
\usepackage{txfonts}
\usepackage{epsfig}
\usepackage{graphicx}
\usepackage{subfig}
\usepackage{caption}
\usepackage{hyperref}

\begin{document} 
   \title{The enigmatic dance of the HD 189733A system: Does the planet accrete onto the star?}
   \subtitle{}
   \author{S. Colombo
          \inst{1}
          \and
          I. Pillitteri\inst{1}
          \and
          A. Petralia\inst{1}
          \and
          S. Orlando \inst{1}
          \and
          G. Micela \inst{1}
          }
   \institute{INAF-Osservatorio Astronomico di Palermo\\
              \email{salvatore.colombo@inaf.it}
             }
   \date{}
% \abstract{}{}{}{}{} 
% 5 {} token are mandatory
  \abstract
   {Several studies suggest that the emission properties of a star can be significantly affected by its interaction with a nearby planet through magnetic fields or interaction between their respective winds. However, the actual observability of these effects remains a subject of debate. An illustrative example is the HD 189733A system, where certain characteristics of its emissions have been interpreted as indicative of ongoing interactions between the star and its associated planet. Other studies attribute these characteristics to the coronal activity of the star.}
   {In this study, we aimed to investigate whether the observed stellar X-ray flare events, which appear to be in phase with the planetary period in the HD 189733A system, could be attributed to the accretion of the planetary wind onto the stellar surface or resulted from an interaction between the planetary and stellar winds.}
   {We developed a 3D magnetohydrodynamic model with the PLUTO code that describes the system HD 189733A , including the central host star and its hot Jupiter, along with their respective winds. The effects of gravity and the magnetic fields of both star and planet are taken into account.}
   {Our analysis reveals that, in the cases examined in this study, the accretion scenario is only viable when the stellar magnetic field strength is at 5 G and the planetary magnetic field strength is at 1 G. In this scenario, the Rayleigh-Taylor instabilities lead to the formation of an accretion column connecting the star to the planet. Once formed the accretion column remains stable for the duration of the simulation.  
   The accretion column produces an accretion rates of the orders of $10^{12}$ g/s and shows an average density of about $10^7 $cm$^{-3}$. 
   In the other case explored, the accretion column does not form because the Rayleigh-Taylor instability is suppressed by the stronger magnetic field intensities assumed for both the star and the planet.
   We synthesized the emission resulting from the shocked planetary wind and, we found that the total X-ray emission ranges from $5\times10^{23}$ to $10^{24}$ erg/s.
   }
   {In the case of accretion, the emission originating from the hot spot cannot be distinguished from the coronal activity. Also, the interaction between the planetary and stellar winds cannot be responsible for the X-ray emission, as the total emission produced is about 4 orders of magnitude lower than the average X-ray luminosity of the star. }
   \maketitle
   
%
%-------------------------------------------------------------------
\section{Introduction}
Hot Jupiters are massive (0.5 to 10 Jupiter masses) gaseous planets orbiting very close (semi-major axis $\le 0.1$ AU) to their host stars. Due to their mass and proximity to the central star, hot Jupiters are the perfect natural laboratories to study star-planet interactions (SPIs). 

Typically, hot Jupiters' atmospheres are heated up by the radiation arising from the central star \citep{Burrows2000,OwenJackson2012,Buzasi2013,Lanza2013}.
In some cases, the heated planetary gas has enough thermal energy to escape from the gravitational field of the planet \citep{Lammer2003}. This phenomenon is called photo-evaporation.
Evidence of these outflows have been reported for example for the systems HD 189733b and HD 209458b \citep{Vidal-Madjar2004,Ehrenreich2008,Lecavelier2010,Lecavelier2012,Bourrier2013}. The rate of photo-evaporation depends on the amount of the emitted flux by the star and of the density of the planet. 

In general, the star generates a wind whose physical characteristics (density and velocity) depend on the star's age \citep{Matt2008,Cohen2014}. The stellar wind is the medium where the planetary outflow expands and interacts. This interaction could produce observable signatures that may result in enhanced stellar activity. \cite{Shkolnik2003,Shkolnik2005,Shkolnik2008} and \cite{Walker2008} analyzed some systems with hot Jupiters and reported evidence of chromospheric activity in phase with the planetary period.
Further, X-ray observations also suggest a correlation between the planetary phase and the stellar activity on HD 189733A \citep{Pillitteri2015,Pillitteri2010,Pillitteri2011, Pillitteri2014}. 
HD 189733 is a wide binary system composed of a K1 type star (HD 189733A) and a M star (HD 189733B) \citep{Bakos2006}. HD 189733A hosts a transiting hot Jupiter (HD 189733b) in a 2.2 days orbit \citep{Bouchy2005}. This system represents one of the best targets to study SPI because of its proximity, strong activity, and the presence the transiting planet, which allows the observation of planetary photoevaporation. HD 189733A has an X-ray luminosity of order of $10^{28}$ erg s$^{-1}$ and a corona brighter than the solar one, with a mean temperature around 5 MK. The age of HD 189733 A derived from its level of activity and X-ray emission appears to be about 1 Gyr and notably younger than the age of the M companion HD 189733B. This discrepancy has been explained with the exchange of angular momentum between the star and the planet \citep{Pillitteri2010,Poppenhaeger2014}. 

\cite{Pillitteri2010,Pillitteri2011,Pillitteri2014} observed at several epochs X-ray flares at planetary phase range $\phi \approx 0.54-0.52$, and $0.64$. Moreover, the flares appear to be colder than the stellar corona (kT$<1.0$keV or T$<12$MK), suggesting that they may have a different origin than stellar activity \citep{Pillitteri2015}.
It was inferred that the planetary magnetosphere may affect the coronal structure of the star \citep{Pillitteri2010}. In particular, \cite{Pillitteri2010} suggested that the planetary magnetosphere may perturb the stellar magnetic field twisting the magnetic field lines and, as a result, producing energetic flares by magnetic reconnection \citep{Pillitteri2010}. This hypothesis is well supported also by theoretical models \citep{Shkolnik2008,Lanza2009}. 
\cite{Pillitteri2014} inferred the flare properties from the oscillation timescales and concluded that the flaring loop generated after the transit is long $\approx25\%$ of the distance between the star and the planet.
Moreover, the density of the emitting region is higher than that of the solar corona as observed from OVII and NeIX spectral lines \citep{Pillitteri2011,Pillitteri2014}. 
A parallel interpretation of the origin of the X-ray emission is proposed by \cite{Pillitteri2015}.
They collected data at far ultraviolet (FUV) wavelengths with the Hubble Space Telescope Cosmic Origin Spectrograph and observed two flaring episodes at $\phi \approx 0.525$ and $\phi \approx 0.588$ that appear to coincide with the X-ray activity observed previously.
They suggest that part of the evaporating material from the planet is intercepted by the stellar gravitational field and accretes  onto the star, forming an accretion column. The impact of the supersonic accretion column with the stellar surface generates a hotspot which is responsible for the enhanced UV and X-ray emission. 

This idea stems from the work of \cite{Matsakos2015}, where an SPI magnetohydrodynamic (MHD) model was developed.
They did not focus specifically on the HD 189733 system, they developed an idealised star-planet system which they used to describe the different kinds of SPIs.
Depending on the physical conditions of the system, various type of interaction may come into play. 
The planetary outflow might be forced by a strong stellar wind in a cometary tail which trails the planet along the orbit \citep{Matsakos2015} and produces observable signatures in phase with planetary period \citep{Mura2011,Kulow2014}. 
Alternatively, the planetary wind might produce a bow shock during the expansion in the stellar wind \citep{Vidotto2010,Vidotto2011a,Vidotto2011b,Llama2011,Llama2013}. Then in some cases, it might be captured by the gravitational field of the central star. When this occurs, the stellar wind is forced to spiral down into the star. The stellar magnetic field funnels the falling gas onto an accretion column-like structure. The impact of this accretion flow with the stellar surface might produce a hotspot that might generate an observable excess of UV and X-ray radiation \citep{Lanza2013,Matsakos2015,Pillitteri2015}. The hotspot orbits in sync with the planetary period but experiences a phase shift of $\approx 90 \deg$. 

Despite the observational evidence supported by theoretical works, there are other explanations for the X-ray activity in HD 189733. In fact, \cite{Route2019} claim that there is no statistical evidence for a bright hotspot synchronized to the planetary orbital period. 
Using Kolmogorov-Smirnov and Lomb-Scargle periodogram analyses, they did not find evidence for persistent hotspots that have locations synchronized to the planetary orbital period.
They suggest that the bright regions that persist for a few rotational periods are entirely consistent with the normal evolution of active regions on stars.  
In addition, they claim that the accretion rate on HD 189733a is at least two orders  of magnitude less than the typical values observed in CTTSs. According to \cite{Route2019} this small accretion rate would produce an undetectable emission.  

In this paper, we analyzed the dynamic of HD 189733, and the observability of the UV and X-ray band of the SPI. Our questions are: a) Is the star accreting material from the planetary winds? b) Is this kind of phenomenon producing visible signatures like those observed \citep{Pillitteri2010,Pillitteri2011,Pillitteri2014,Pillitteri2015}?
To address these questions we developed an MHD model that describes the star-planet system in HD 189733A. We then synthesized high energy emissions from the results of the simulation. 

The paper is structured as follows: In Sect. \ref{sect:model} we describe the MHD model used in this work, in Sect. \ref{results} we discuss the results from the simulation including also the results from the synthesis of emission and the implication on the observability of this kind of SPI. Finally, in Sect. \ref{conclusions} we draw our conclusions. 

%--------------------------------------------------------------------
%
%--------------------------------------------------------------------
\section{Model}\label{sect:model}
Star-planet systems with evaporating hot Jupiters can be fully described by making use of hydrodynamic or magnetohydrodynamic (MHD) models, as discussed in \cite{Matsakos2015} (see also reference therein).  In this work, we adopted a MHD model analogous to the one described in \cite{Matsakos2015}, but in this case, as previously discussed, the setup mimicking the system HD 189733A. 

HD 189733 is a wide binary system located at a distance of $\approx 19.3$ pc from Earth. The two stars of the system are separated by an average distance of $D_s \sim 220$ AU. HD 189733A is a K1.5V type star with mass $M_{s}=0.805$M$_\sun$ radius $R_{s}=0.76R_{\sun}$, and a rotational period of $\approx11.95$d \citep{Henry2008}. This star hosts a hot Jupiter with mass $M_{p}=1.13$M$_J$ that orbits at a distance of $D_p \sim 0.031$AU with a period of P$_p=2.219$ days \citep{Bouchy2005}.
Winds originate from both the star and the planet and propagate through the interplanetary medium.

In the following we will focus on in the planetary system of HD 189733A. Due to the fact that $D_p \ll D_s$ we assumed that the effects of the companion star are negligible in the dynamics of the star-planet system.

\subsection{Equations}
We adopted a spherical coordinate system ($R$,$\theta$,$\phi$) centered at the center of the star.
Since $M_{s}\gg M_{p}$ we consider the center of the star as the center of mass of the entire system. The planet orbits around the star in the $\theta = \pi/2 $ plane. The system is described in the reference frame co-rotating with the planet. In this reference frame, the model solves the MHD equations of conservation of mass, momentum and energy and also the equation of induction for the magnetic field and the equation of entropy. 
\begin{gather}
	\frac{\partial}{\partial t}\rho + \nabla \cdot  \rho\vec{u} =0\\	
	\frac{\partial}{\partial t}\rho\vec{u}+\nabla \cdot (\rho\vec{u}\vec{u} - \vec{B}\vec{B}+ \vec{I}p_t) = \rho\vec{g}\\		
    \frac{\partial}{\partial t}\rho E+\nabla\cdot[(\rho E+p_t)\vec{u}-\vec{B}(\vec{u}\cdot\vec{B})] = \rho \vec{u}\cdot(\vec{g}+\vec{F_{ext}}) \label{RL} \\
	\frac{\partial}{\partial t}\vec{B}+\nabla\cdot(\vec{u}\vec{B}-\vec{B}\vec{u})=0 \\
	\frac{\partial\rho\sigma}{\partial t} + \nabla \cdot (\rho\sigma\vec{u}) = 0
\label{eq_Q}
\end{gather}
\noindent
where: 
\begin{equation}
p_t= P+\frac{\vec{B} \cdot \vec{B}}{2} , \qquad E = \epsilon + \frac {\vec{u} \cdot \vec{u}}{2}+\frac{\vec{B} \cdot \vec{B}}{2\rho}
\end{equation}
where, $\rho$ is the plasma density, $\vec{u}$ the plasma velocity, $p_t$ the total plasma pressure, $\vec{B}$ the magnetic field, $g$ the gravity acceleration vector, $F_{ext}$ is an inertial force that appears in our non-inertial rotating frame. $F_{ext}=F_{Coriolis}+F_ {centrifugal}$ has a Coriolis and centrifugal components given by: $F_{Coriolis} = 2(\Omega_{fr}\times \vec{v})$ and $F_{centrifugal}=-\left[\Omega_{fr}\times(\Omega_{fr}\times R)\right]$. 
$E$ is the total plasma energy density, $\epsilon$ the thermal energy density and $\sigma= p/\rho^\gamma$ the plasma entropy.
We use the ideal gas law $P = (\gamma -1)/\rho\epsilon$, where the polytropic index $\gamma = 3/2$. 

The calculation was performed using the PLUTO code, a modular Godunov-type code for astrophysical plasmas \citep{Mignone2012}. The code uses parallel computers using the Message Passage Interface (MPI) libraries. The MHD equations are solved using the MHD module available in PLUTO with the Harten-Lax-van Leer Riemann solver. The time evolution is solved using the Hancock method. To follow the magnetic field evolution and to maintain the solenoidal condition we use the eight waves technique \citep{Powell1994,Powell1999}.

\subsection{Initial and Boundary conditions}
To impose the initial conditions we first set up the stellar wind which is generated at the surface of the star and expands in a low density static ambient medium ($\rho = 10^{-18}$g cm$^{-3}$). The wind is generated at the stellar surface, which is described in the region where $R \leq 1.1$R$_\odot$. In particular, for $R < 0.9$R$_\odot$ $\rho$, $p$ and $\vec{v}$ are fixed to generate the stellar wind and $\vec{B}$ is fixed to describe the dipole configuration for the stellar magnetic field. In regions where 0.9$_\odot$<R<1.1$_\odot$, $\rho$, $p$ and $\vec{v}$ are still fixed to generate the wind, but in order to increase the code stability, we set the magnetic field in this region as free to evolve.
We let this setup evolve until it reaches a stationary condition ($t \approx 1$P$_p$). Then we used this evolved setup in which we added the planet and its wind and unperturbed magnetic field as the initial conditions for our simulation.
At $t=0$, the planetary wind is generated from the planetary surface. The planetary wind is prescribed in regions inside $1.5R_p$ where $\rho$, $p$, $\vec{v}$ and $\vec{B}$ are fixed. The stellar and planetary interiors are not involved in the simulation and they are prescribed as internal boundaries. As a result, the physical quantities are fixed and the solution is overwritten at each time step. We approximate the star and the planet as a rotating solid bodies that rotate axes parallel to the z-axis.
The parameters that characterize the star and planetary winds are summarised in Table \ref{table:values}.
\begin{table}[h!]
\centering
\caption{Values of interest of the winds at the stellar and planetary surfaces.} 
\label{table:values}
    \begin{tabular}{c|c|c}
                          & Star            & Planet \\
                            \hline
   Density (g cm$^{-3}$)  &  $4.32\times 10^{-18}$      & $2.16\times10^{-16}$  \\ 
   Temperature (K)        & $10^6$          & $10^4$              \\
   Velocity (cm s$^{-1}$) & $6.5\times10^{7}$ & $6.6\times 10^{6}$ \\
  
    \end{tabular}
\end{table}

The primary objective of this study is to investigate the intricate interplay between the planetary wind and the stellar wind and its potential to generate accretion events onto the star, which consequently produce discernible X-ray signatures. To achieve this, we have opted for an exceptionally high evaporation rate ($\dot{M}\approx 10^{-9}$M$_J$/yr) for the planet while considering a relatively low density stellar wind. 

The initial configuration of the global magnetic field results from the combination of the stellar and planetary magnetic fields. The stellar magnetic field is modelled as a Parker spiral, while the planetary magnetic field is represented by a dipole configuration. We have explored the influence of the magnetic field strength on shaping the dynamics of the stellar wind. This choice allowed us to assess the impact of the magnetic field strength on the SPI.
By incorporating these aspects into our study, we aim to gain deeper insights into the interaction between the planetary wind and the stellar wind, the occurrence of accretion events onto the star, and the resultant observable X-ray signatures.
Table \ref{tab:sims} shows all the different cases explored in this work.

\begin{table}[h!]
    \centering
    \caption{Cases explored in this work.} 
    \label{tab:sims}
    \begin{tabular}{c|c|c}
        Name & $B_s$ (G) & $B_p$ (G) \\
        \hline
        Bs5-Bp1 (Reference case)    &  5 & 1 \\
        Bs5-Bp5                     &  5 & 5 \\
        Bs10-Bp1                    & 10 & 1 \\
        HD                          &  0 & 0 \\ 
    \end{tabular}
\end{table}

The inner boundary in $R$ is defined to describe the stellar surface and to generate the stellar wind. The external boundary in $R$ and the boundaries in $\theta$ are prescribed as gradient equals 0 (outflow condition) for density, pressure and velocity. The magnetic field at the outer boundary is calculated via parabolic extrapolation using the three closest grid cells. The boundaries for $\phi$ are located at $0$ and $2\pi$ for this reason they are assumed to be periodic. 

\subsection{Spatial Grid}
\begin{figure}
\centering
     \includegraphics[width=0.95\hsize]{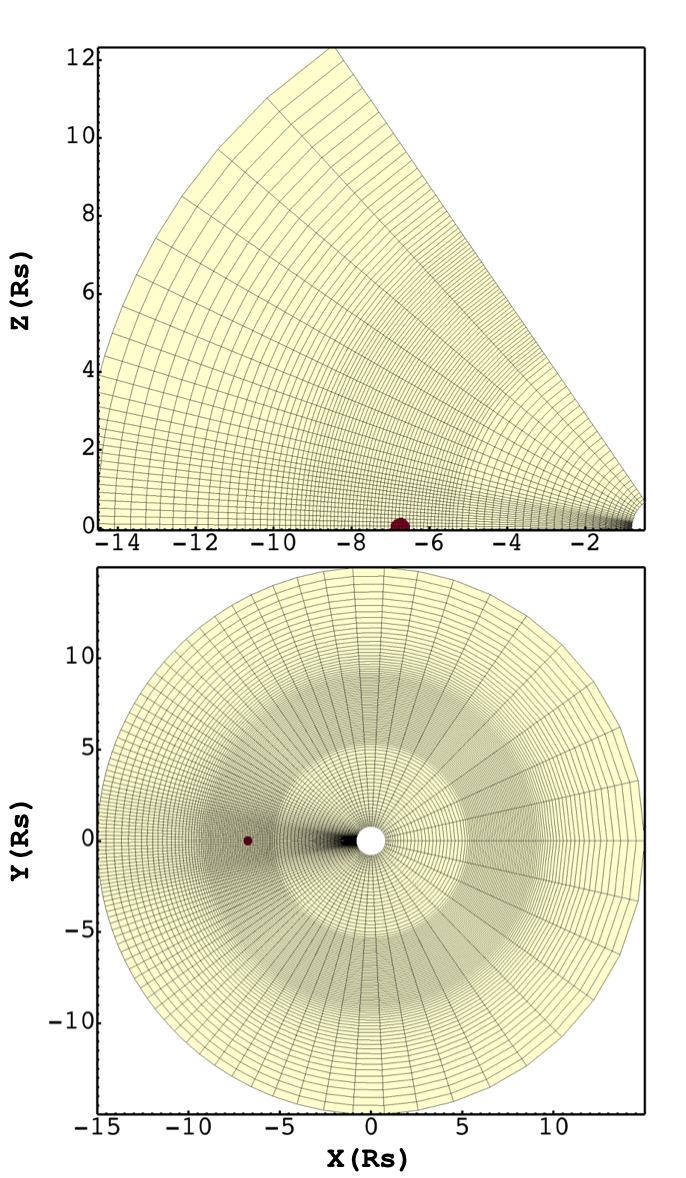}
     \caption{Structure of the numerical grid used in the model. The top panel displays a slice of one quadrant of the $xz$ plane, to enhance clarity and due to symmetry half domain is shown. The bottom panel illustrates a slice in the $xy$ (equatorial) plane. The dark red dot represents the planet. To enhance clarity, the black grid presented is four times coarser than the adopted resolution (i.e., each box contains $4\times 4$ grid points).}\label{fig:Grid}
\end{figure}

Our simulations need a grid that adequately resolves the small-scale planetary wind structures in a huge domain. For this purpose, a uniform grid is too expensive from a numerical point of view. Our strategy follows from \cite{Matsakos2015}. The system is described in a proper corotating frame of reference in which the star and planet are fixed. In this reference frame the planet is located at $(r,\theta,\phi) \equiv (D_p, \pi/2, \pi)$. We, also, make use of a non-uniform spherical grid centered at the center of the star in order to increase the spatial resolution close to the planet and to the stellar surface. The grid is showed in Fig. \ref{fig:Grid}

The radial coordinate $r$ extends from the stellar radius (i.e. $r_{min}=0.805 R_\odot$) to $r_{max}=15R_\odot$ and it is divided in 5 parts: the first ranges from $r=0.805R_\odot$ to $r=2R_\odot$ with 36 points and the resolution of mesh decreases with $r$ in order to have more resolution close to the stellar surface; the second grid ranges from $r=2R_\odot$ to $r=5R_\odot$ with 60 points with a uniform resolution of $0.05 R_\odot$; the third part is composed of 14 points and ranges from $r=5R_\odot$ to $r=5.5R_\odot$; the fourth part of the grid that ranges from $r=5.5R_\odot$ to $r=9.0R_\odot$ is uniform with 150 points that correspond to a resolution of $0.03 R_\odot$; the fifth part  of the grid ranges up to $r=15R_\odot$ with 100 points and the resolution decreases with $r$ 
(See Fig.\ref{fig:Grid}).

The $\theta$ coordinate ranges between $\theta_{min} =35^{\circ}$ and $\theta_{max} =145^{\circ}$ and it is composed of three parts: a central uniform and more resolved grid ranging with 100 points from $\theta_{min}=80^{\circ}$ to $\theta_{max}=100^{\circ}$ and two grids with 40 points from $\theta_{min}=35^{\circ}$ to $\theta_{max}=80^{\circ}$ and $\theta_{min}=100^{\circ}$ to $\theta_{max}=145^{\circ}$ respectively
(See Fig.\ref{fig:Grid} (b)).

An analogous scheme is adopted for the $\phi$ coordinate. In particular, the $\phi$ coordinate is divided into three grids: a central uniform and more resolved grid with 100 points from $\phi_{min}=170^{\circ}$ to $\phi_{min}=190^{\circ}$ and two with 150 points each ranging from $\phi_{min}=0^{\circ}$ to $\phi_{min}=170^{\circ}$ and $\phi_{min}=190^{\circ}$ to $\phi_{min}=360^{\circ}$. This grid ensures the highest resolution close to the planet and in the planetary plane (See Fig.\ref{fig:Grid} (a)).

%(inserire una figura per far vedere i livelli?)

\subsection{Synthesis of Emission\label{sect:em}}
The code output are the values of density, pressure, components of velocity and magnetic field for each cell in the grid and 2 passive tracer used to identify the planetary wind and the stellar wind. 
From the model results we synthesised the emission in the X-ray band [0.1-10keV]. 
From the 3D domain, we calculated the emission measure from the $j_{th}$ cell, with density $\rho_j$ and volume $dV_j$ as $EM_j =\rho_j^2 dV_j$. For each cell we synthesized the emerging spectrum by multiplying the EM by the spectrum per unit of EM obtained from the CHIANTI atomic lines database \citep{Landi2013}, assuming solar abundances. 
To obtain the total X-ray emission we integrated the spectra in the band [0.1-10KeV] obtained for each cell in the whole spatial domain. Here we are assuming that the material in the domain is optically thin.
%--------------------------------------------------------------------
%
%-------------------------------------------------------------------
\section{Results}\label{results}
In this section, we present the results of the numerical simulations and the synthesis of the emission for the 4 cases.
In the following, we consider the simulation with the lowest value of $B_s$ and $B_p$ as reference case. The other simulations are named as in Tab. \ref{tab:sims}. The simulations do have the same duration, in fact each run was stopped as soon as a stationary regime is reached.
 
\subsection{Dynamics}
\begin{figure*}[!h]
\centering
	\subfloat[\label{fig:RefD10} ]{
        \includegraphics[width=0.45\hsize]{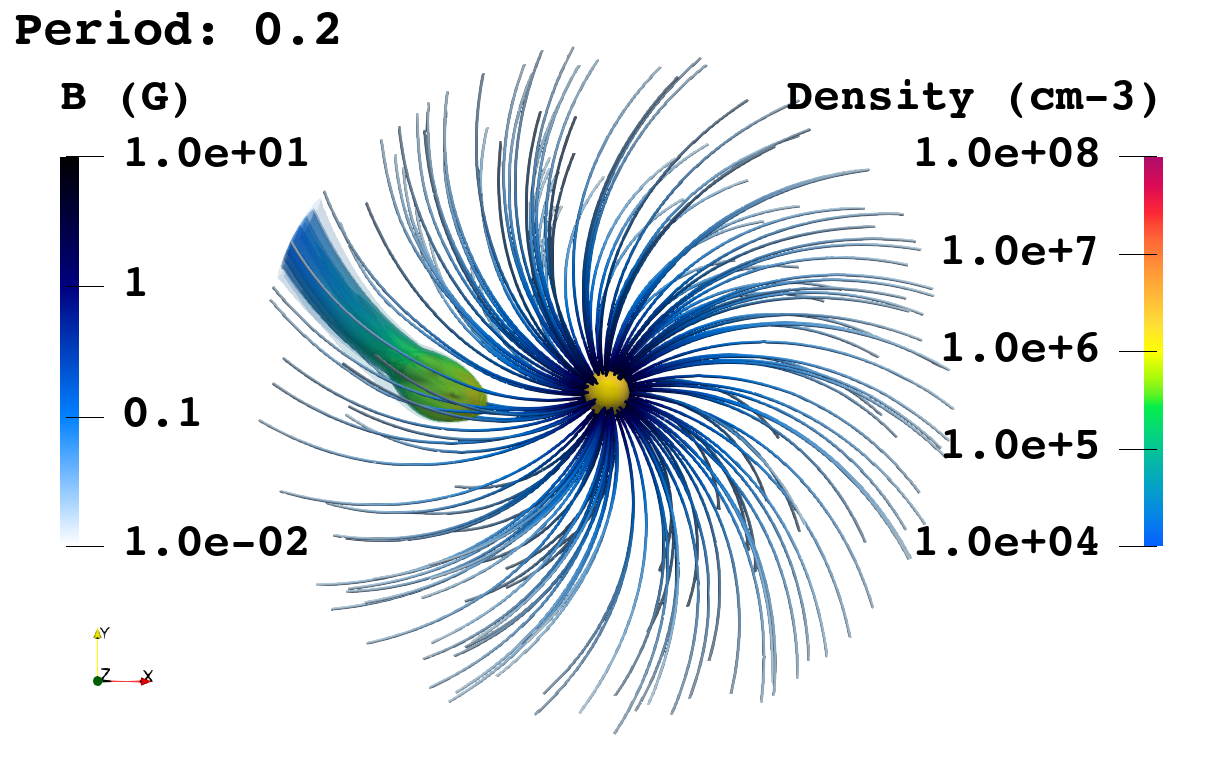}
    }
	\subfloat[ \label{fig:RefD20}]{
        \includegraphics[width=0.45\hsize]{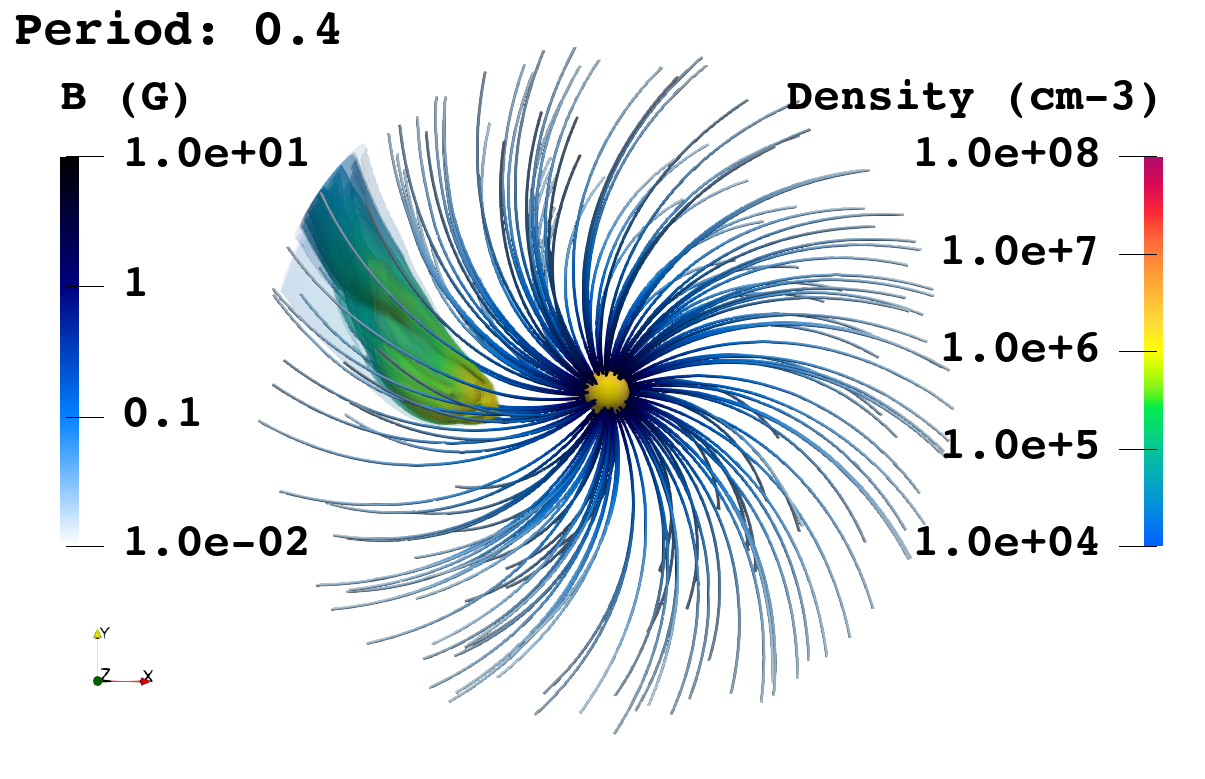}
    }
    
    \subfloat[\label{fig:RefD40}]{
        \includegraphics[width=0.45\hsize]{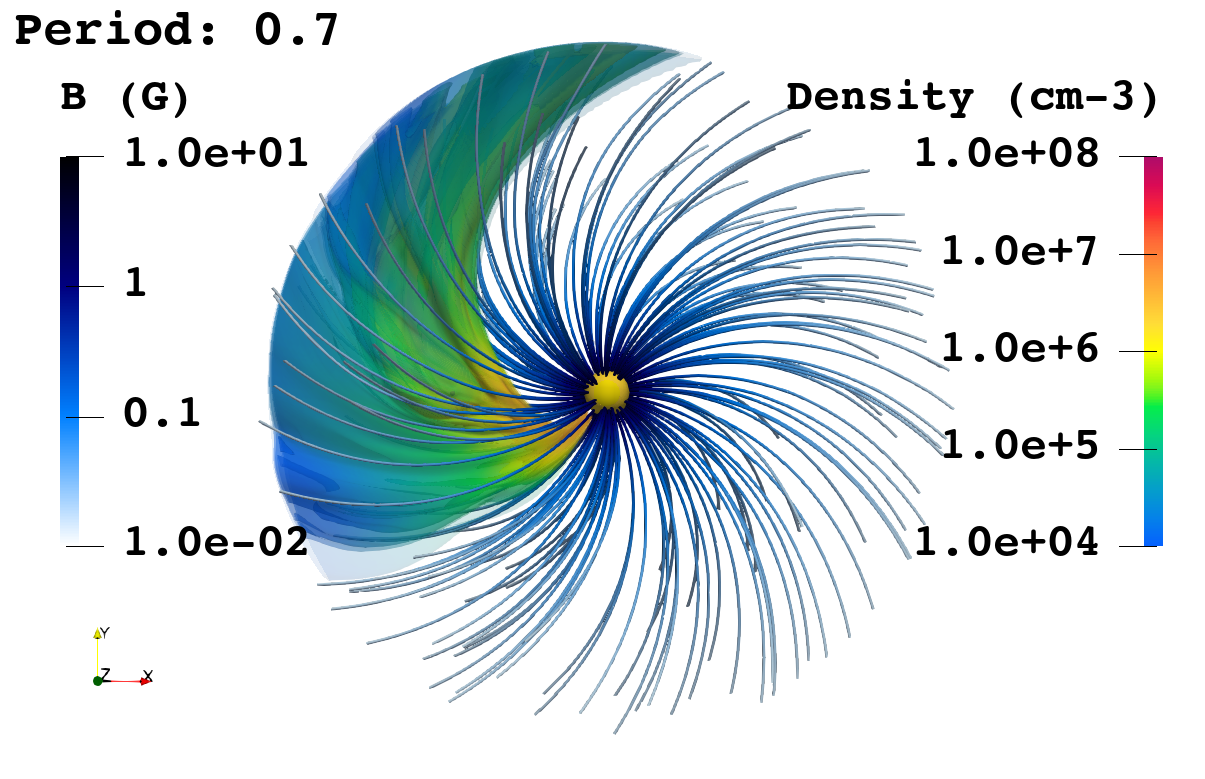}
    }
\subfloat[\label{fig:RefD90} ]{
        \includegraphics[width=0.45\hsize]{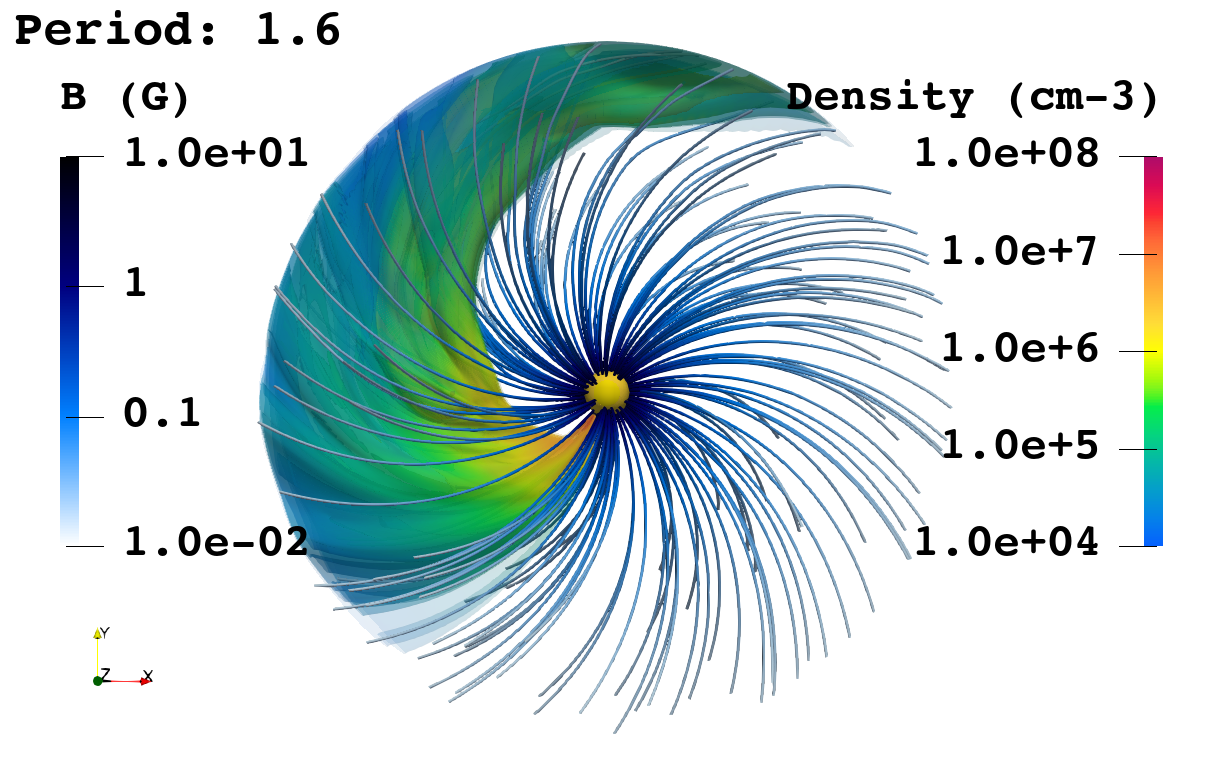}
    }
    
	\subfloat[ \label{fig:RefD120}]{
        \includegraphics[width=0.45\hsize]{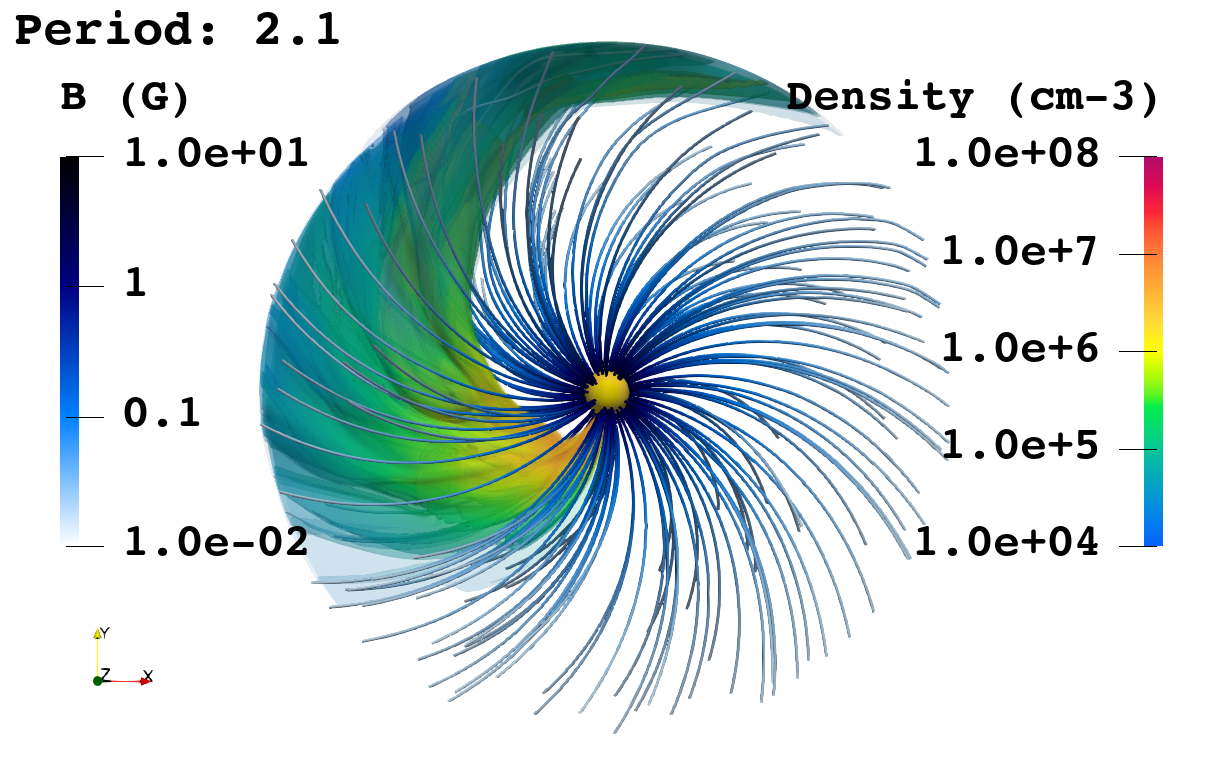}
    }
    \subfloat[\label{fig:RefD160}]{
        \includegraphics[width=0.45\hsize]{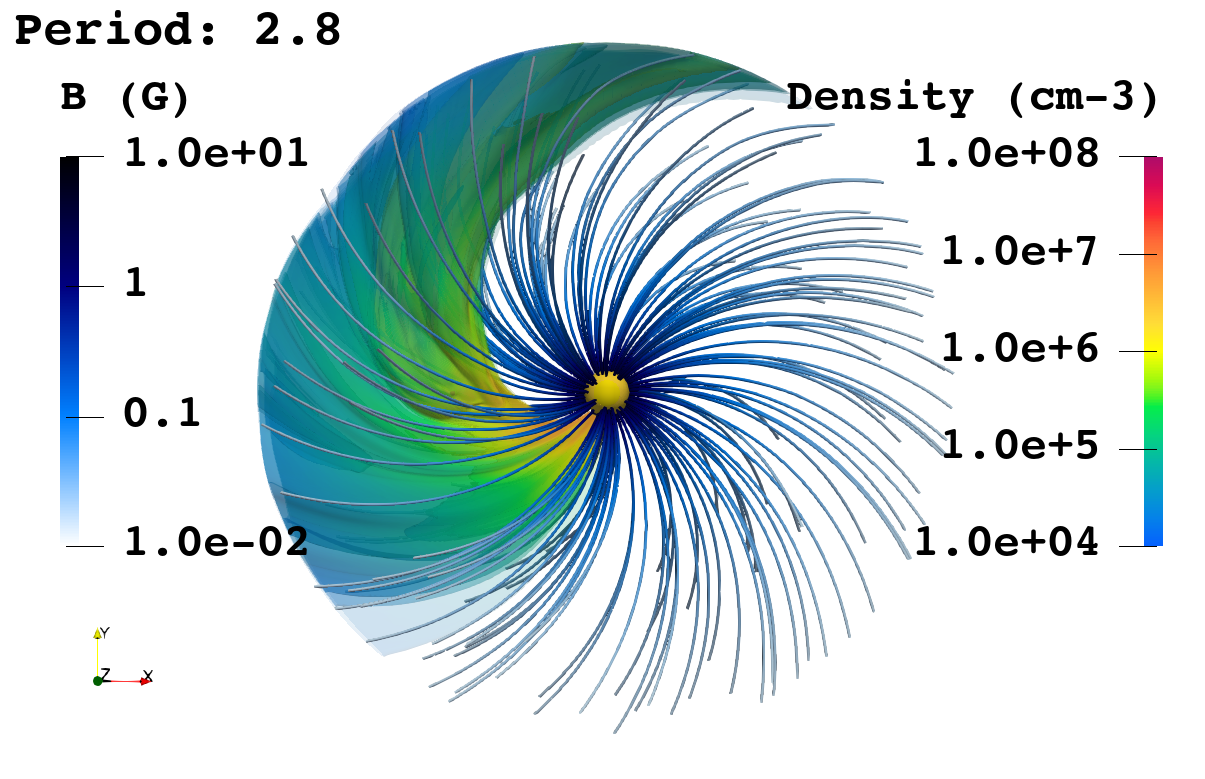}
    }
    \caption{Pole on views of the evolution of density for the Bs5-Bp1 case. The different panels show the density of the planetary wind in a log colour scale (right color bar) at different times as indicated in the top left corner of each image.  The yellow sphere at the centre represents the central star. The planet is located at the left of the star and it is embedded in the planetary wind. The blue-to-white lines represent the magnetic field lines (left color bar).}
    \label{Fig:RefDensEvo}
\end{figure*}

\begin{figure*}
\centering
	\subfloat[\label{fig:RefT10} ]{
        \includegraphics[width=0.45\hsize]{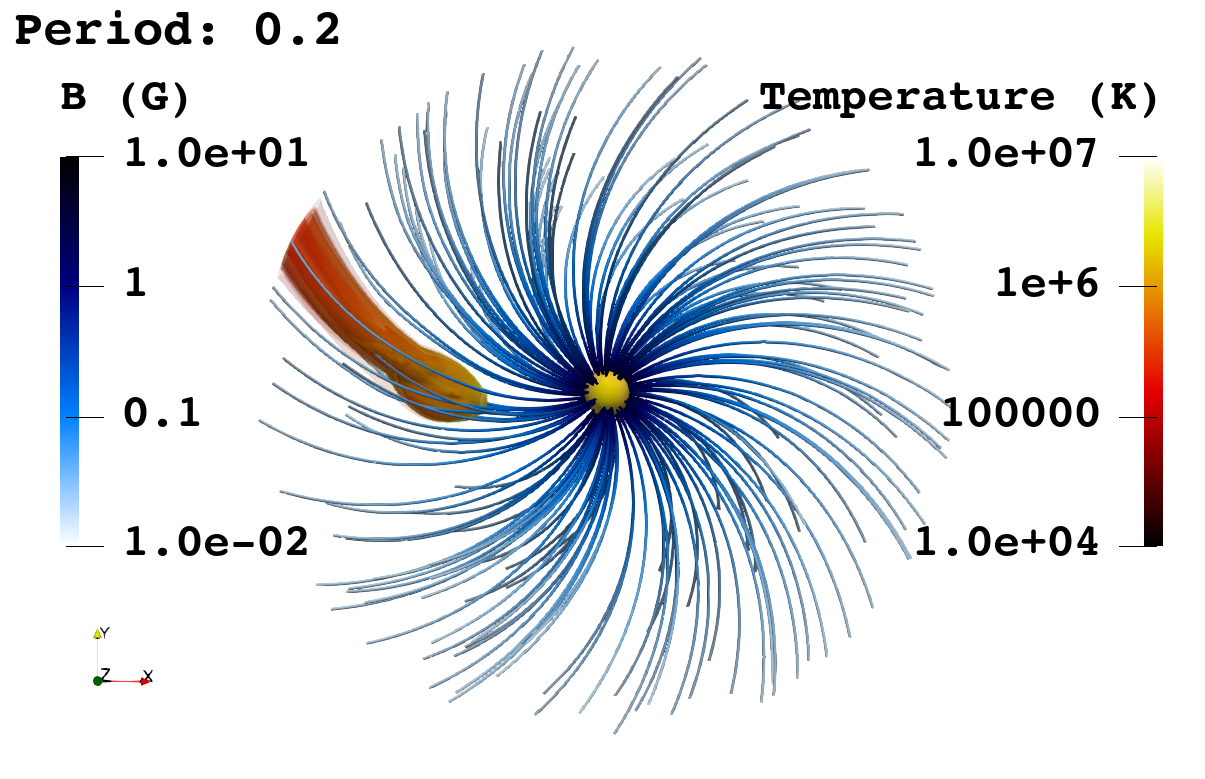}
    }
	\subfloat[ \label{fig:RefT20}]{
        \includegraphics[width=0.45\hsize]{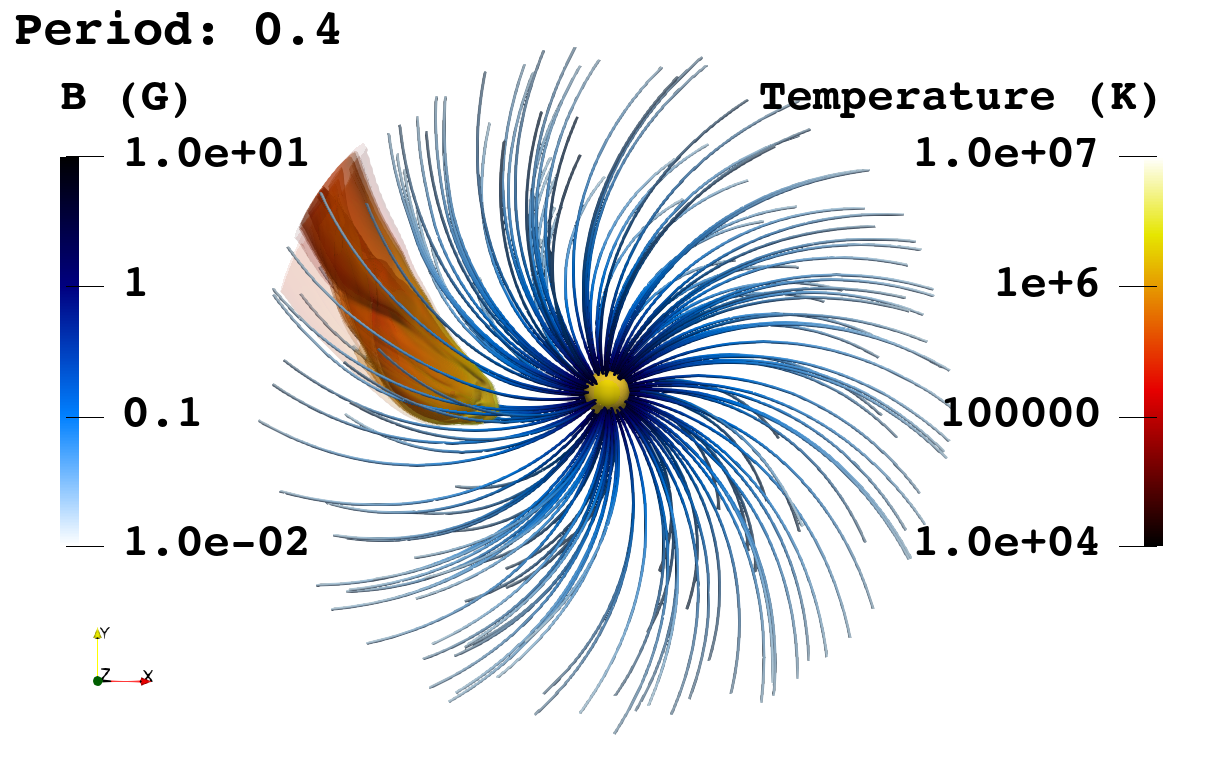}
    }
    
    \subfloat[\label{fig:RefT40}]{
        \includegraphics[width=0.45\hsize]{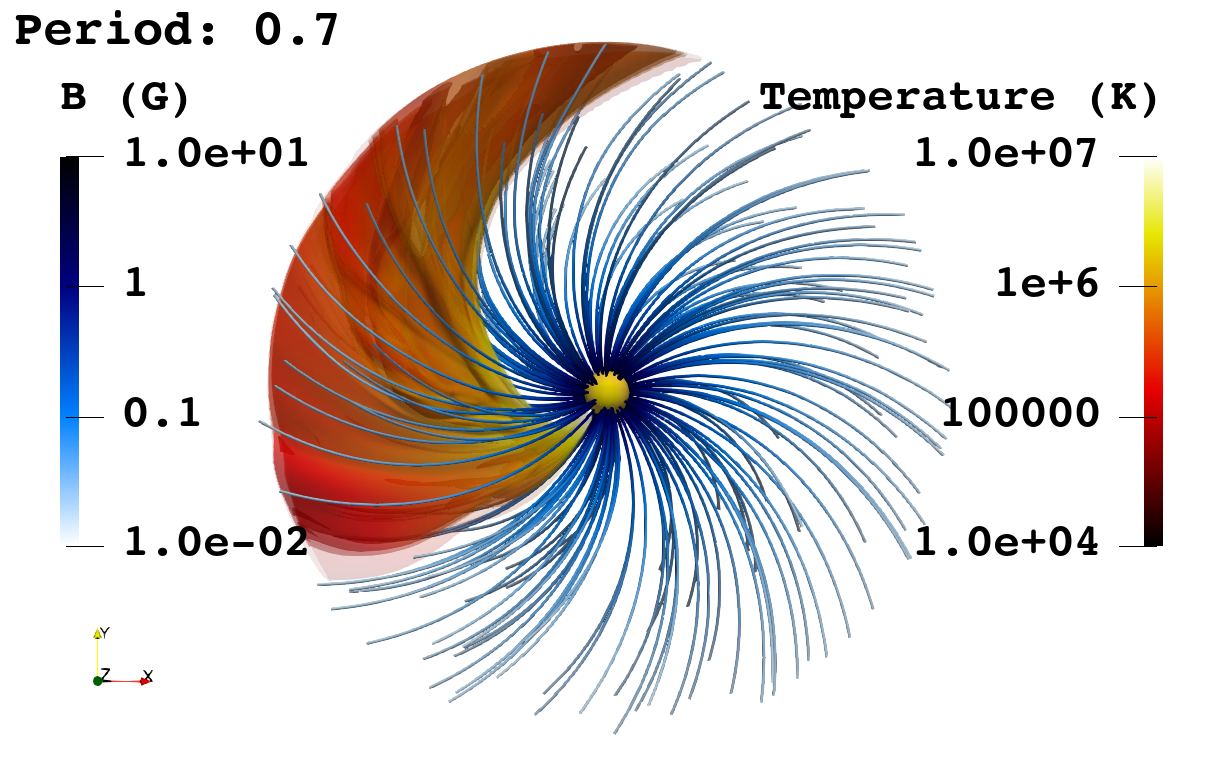}
    }
    \subfloat[\label{fig:RefT90} ]{
        \includegraphics[width=0.45\hsize]{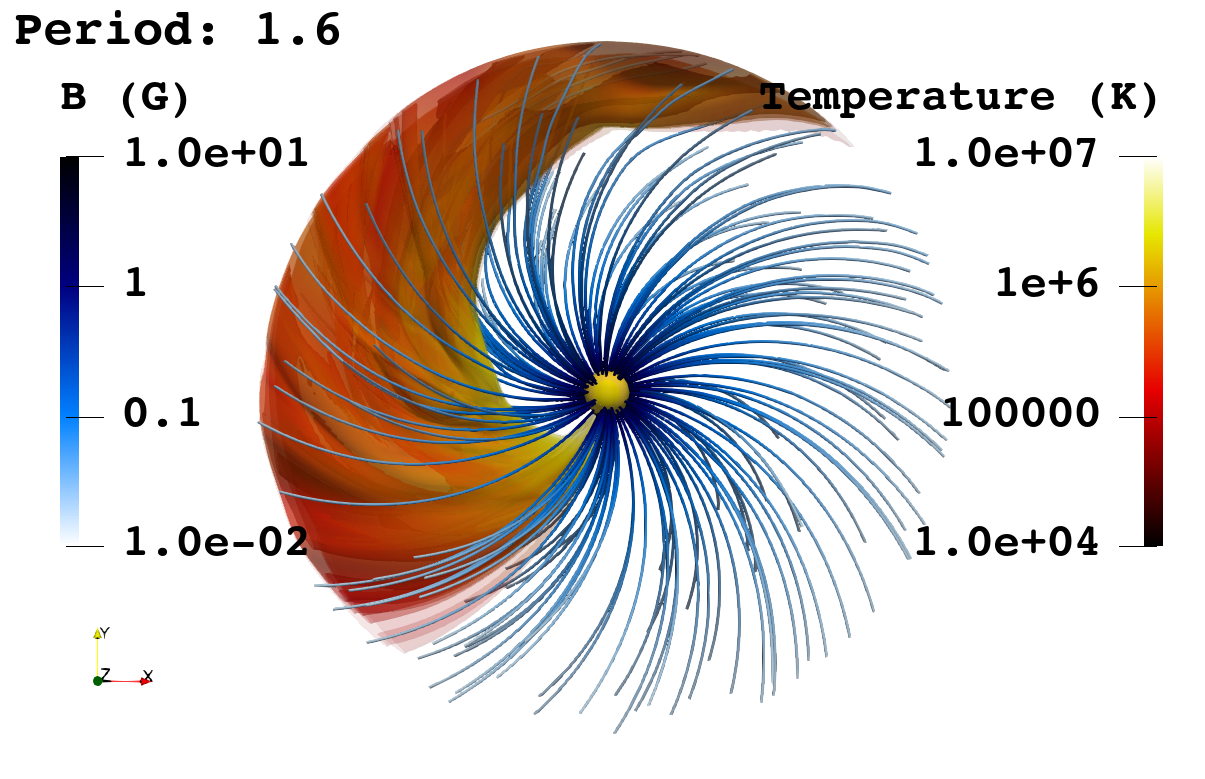}
    }
    
	\subfloat[ \label{fig:RefT120}]{
        \includegraphics[width=0.45\hsize]{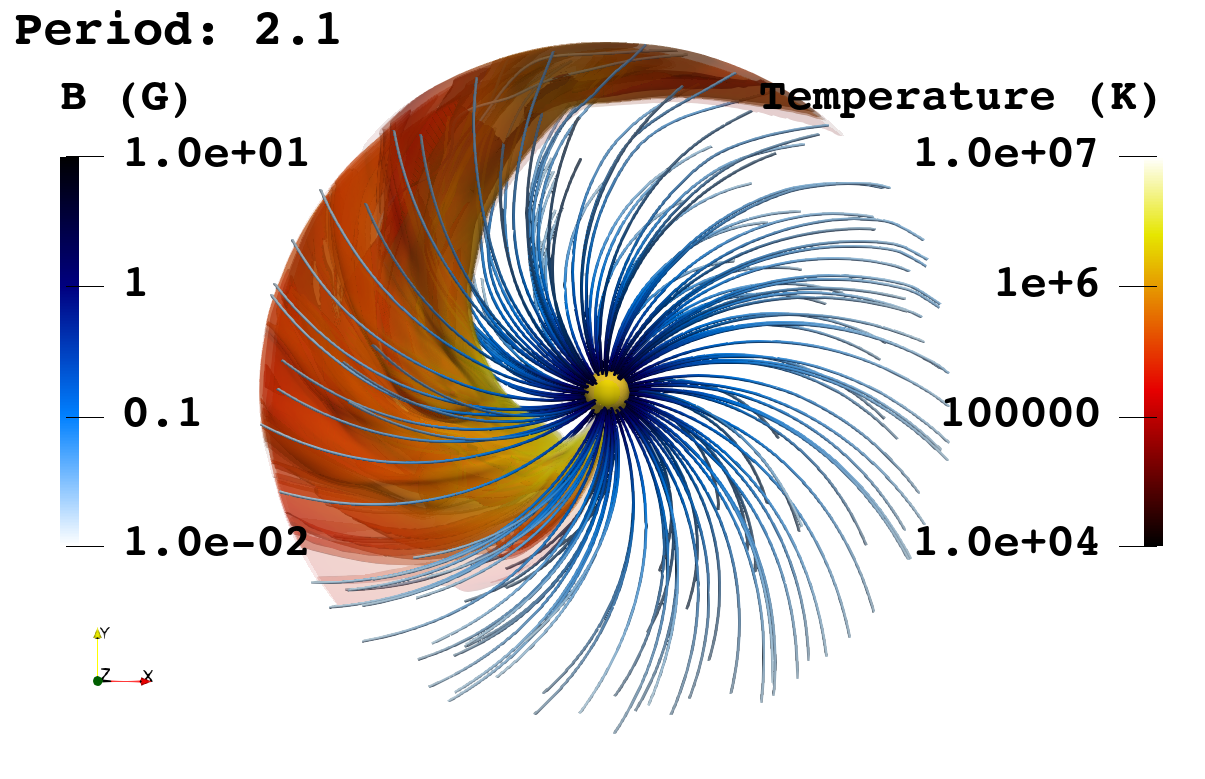}
    }
    \subfloat[\label{fig:RefT160}]{
        \includegraphics[width=0.45\hsize]{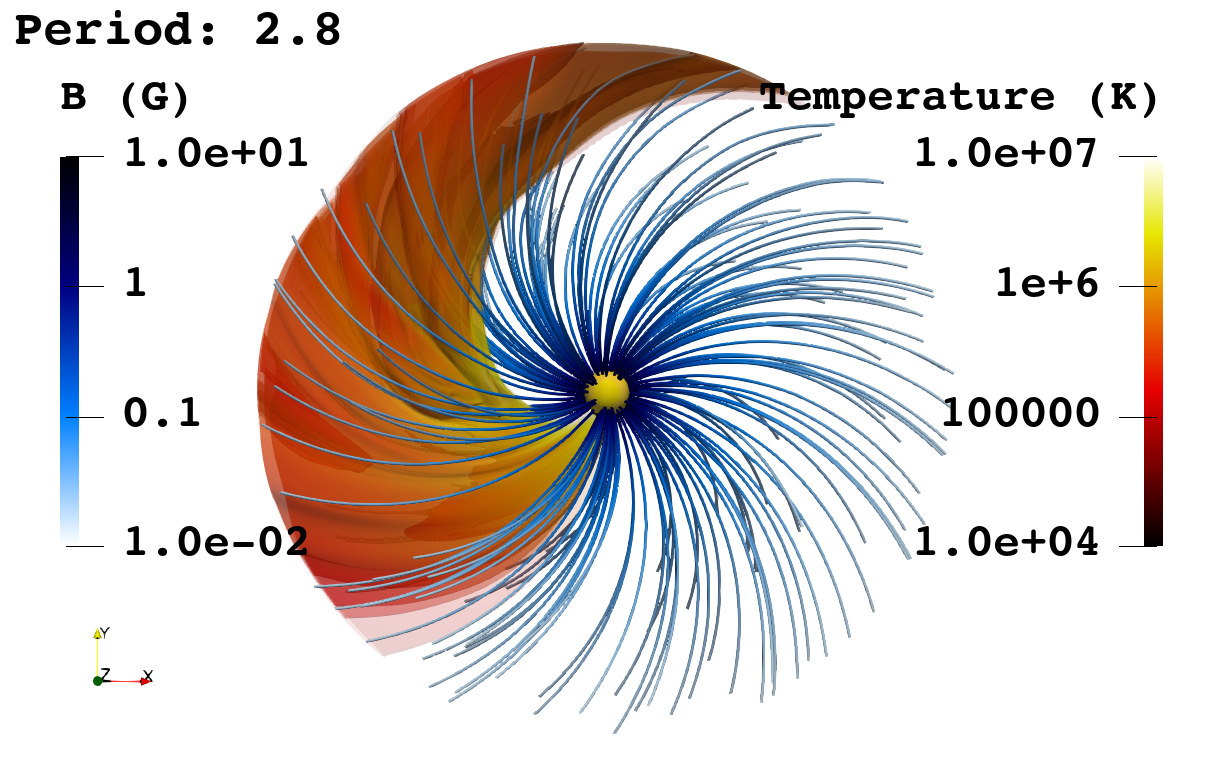}
    }
\caption{As Fig.\ref{Fig:RefDensEvo} but for temperature.}
\label{Fig:RefTempxEvo}
\end{figure*}

\begin{figure}
\centering
	\subfloat[\label{fig:RefB10} ]{
        \includegraphics[width=0.9\hsize]{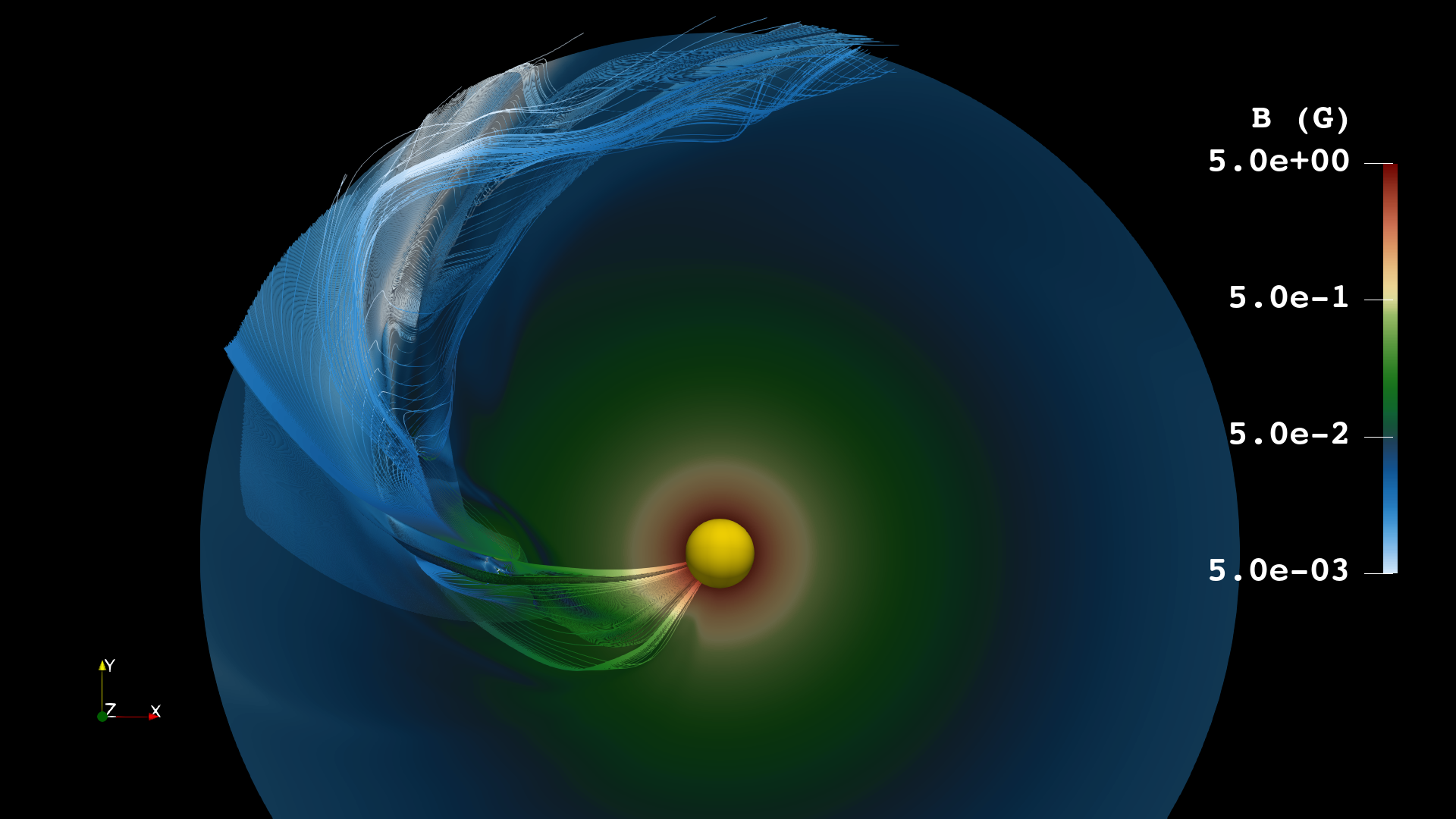}
    }
    
	\subfloat[ \label{fig:RefB20}]{
        \includegraphics[width=0.9\hsize]{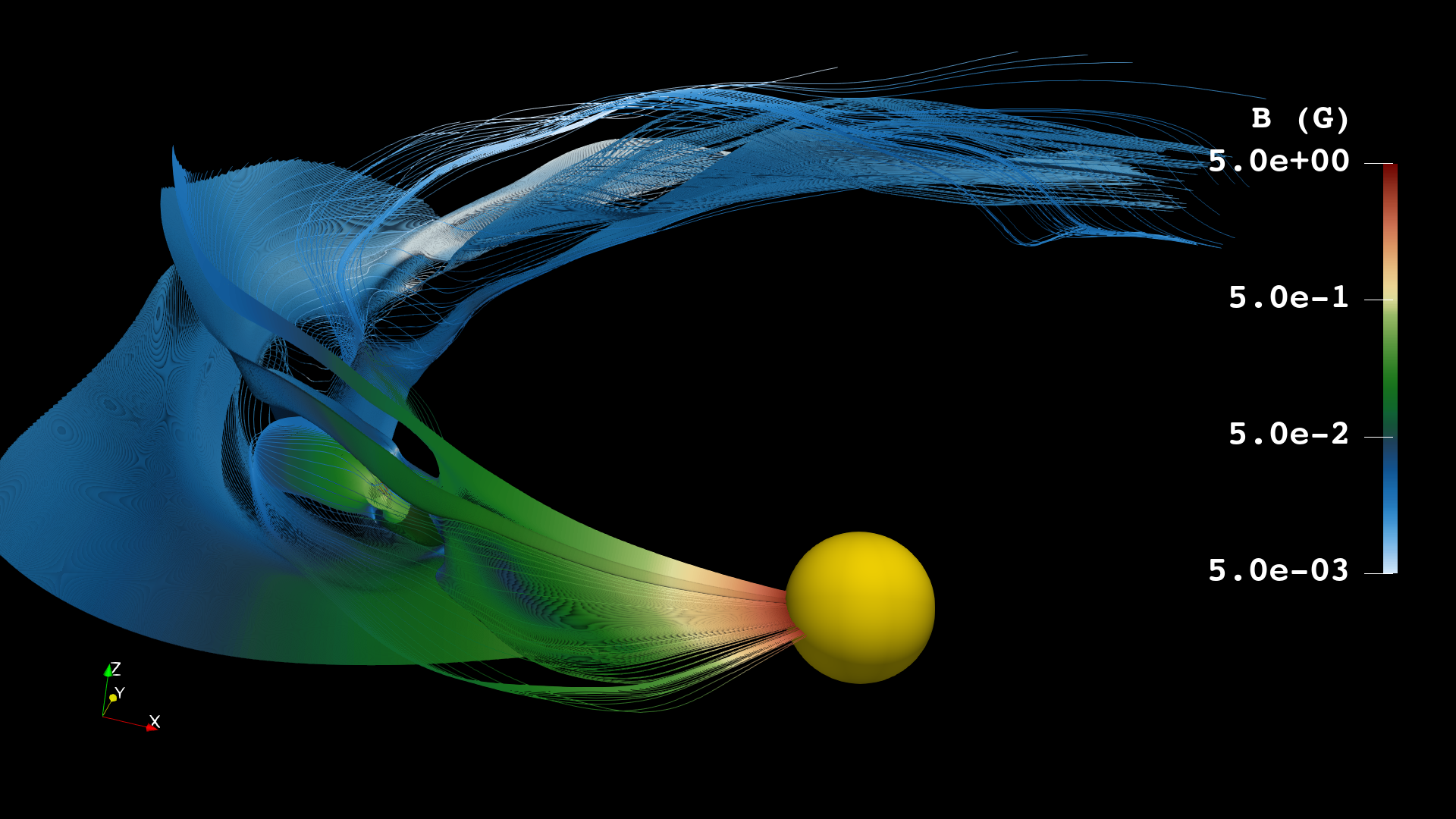}
    }
    
    \subfloat[\label{fig:RefB40}]{
        \includegraphics[width=0.9\hsize]{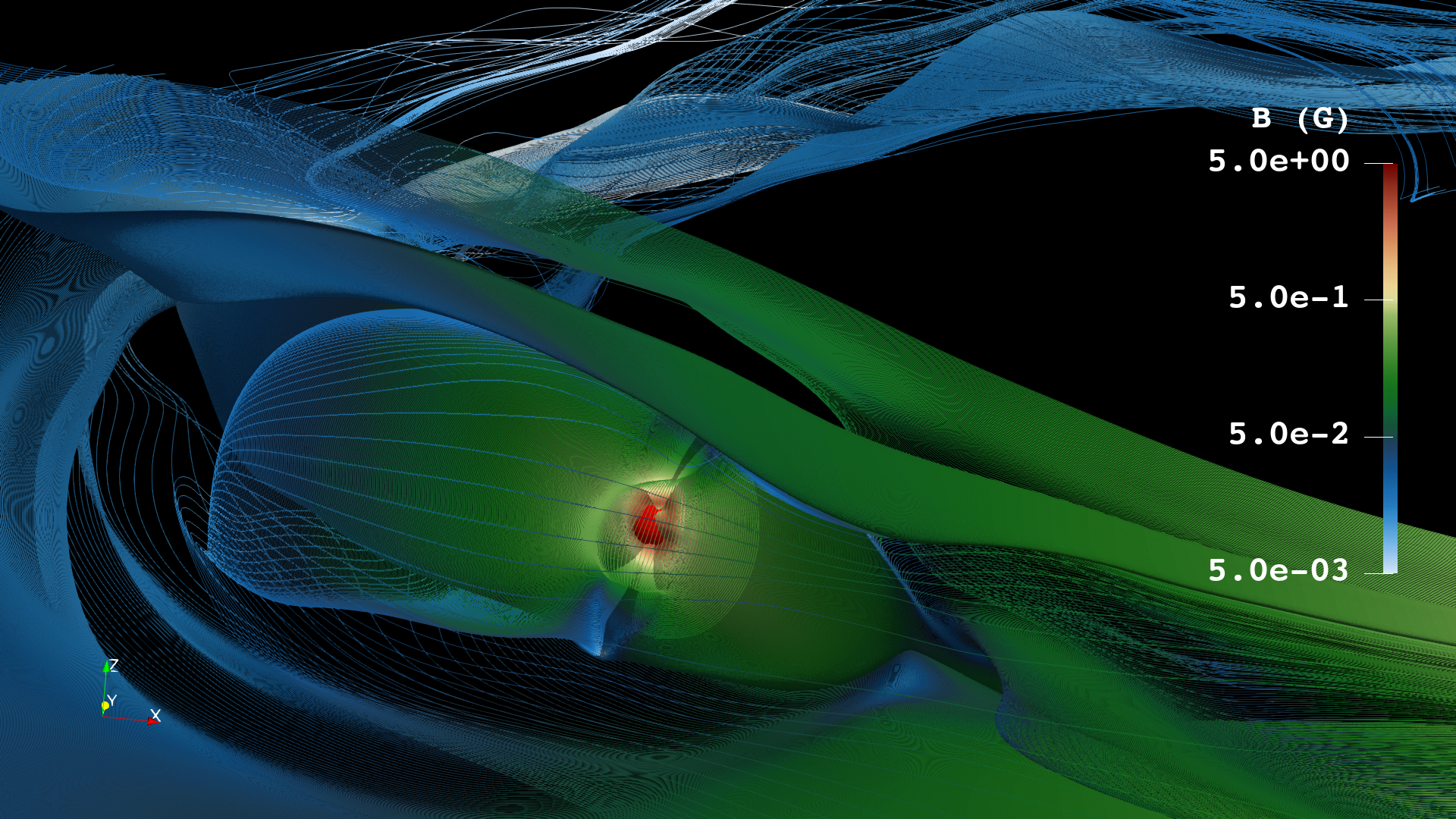}
    }
\caption{Magnetic field configuration for the system. Top panel: pole on view of the system, the equatorial surface shows the magnetic field intensity.Middle panel: view of the magnetic field lines. Bottom panel: close up view of the planetary region. The red sphere is the planet, the yellow sphere is the star. A 3D interactive graphics is available at this link: \url{https://skfb.ly/oOzVG}.}
\label{Fig:RefBEvo}
\end{figure}

Here, we focus on the dynamics of the planetary wind during the simulations. Each simulation was interrupted once a steady state was reached (the configuration of the system do not change for at least half planetary orbit).
Fig. \ref{Fig:RefDensEvo}, Fig. \ref{Fig:RefTempxEvo} and Fig. \ref{Fig:RefBEvo} show the evolution of the planetary winds during the orbits for the reference case (Bs5-Bp1), while Fig. \ref{Fig:RefBEvo} shows the evolution of the magnetic field (stellar+planetary). Analogous two movies with the full evolution of the system are present as additional files. In this scenario, the evolution is similar to the type III case in \cite{Matsakos2015}. 

The planetary wind evaporates from the planet but is confined by the magnetic field and its expansion is favoured by the magnetic field along the planetary orbit (see Fig. \ref{fig:RefD10} and \ref{fig:RefD20}). During the expansion, Rayleigh-Taylor instabilities develop at the contact surface between the planetary and stellar winds. 
The planetary wind is pushed out of equilibrium by the Rayleigh-Taylor instabilities and due to the impact with the stellar wind loose angular momentum triggering the formation of an accretion column of planetary material into the star. The plasma that loses angular momentum is caught by the stellar gravitational field and funnelled by the stellar magnetic field giving rise to a structure akin to an accretion column that links the planet to the star. 
The stellar magnetic field drives the plasma into the accretion column and shapes the topology of the accretion column. The material impacts the stellar surface ahead of the planet at about $45^\circ-60^\circ$ (see Fig. \ref{fig:RefD160}) which is similar to what is suggested by \cite{Pillitteri2015}.
Another part of the planetary wind is pushed away by the stellar wind and spirals away from the system forming a cometary tail structure. In this region, the interaction with the plasma perturbs the stellar magnetic field and generates regions in which reconnection phenomena may occur. 
Furthermore, due to the collision with the stellar wind a front shock at $T\approx10^6$K develops and heats up the wind at temperatures between $T\approx10^5$K and $10^6$K. (Fig. \ref{Fig:RefTempxEvo}).

The planetary wind strongly interacts with the magnetic field, the details of this interaction are shown in Fig. \ref{Fig:RefBEvo}. The result of the interaction between the planetary and stellar winds is a complex magnetic field configuration. In particular, there are regions between the star and the planet where the magnetic field lines twist and The magnetic field structure is notably complex in the region between the planet and the star. In this area, magnetic field lines twist, and regions with opposite-polarity magnetic fields occur. Reconnection events may take place in these regions.(See Fig.\ref{Fig:RefBEvo} c, e and f). However, our models are unable to describe magnetic reconnection events. because they lack  resistivity effects and  the high spatial resolution required to  accurately describe these events. Nevertheless, the study of reconnection events is beyond the scope of the present paper and may be the primary focus of future studies.

\begin{figure*}
\centering
	\subfloat[\label{fig:RefT10} ]{
        \includegraphics[width=0.328\hsize]{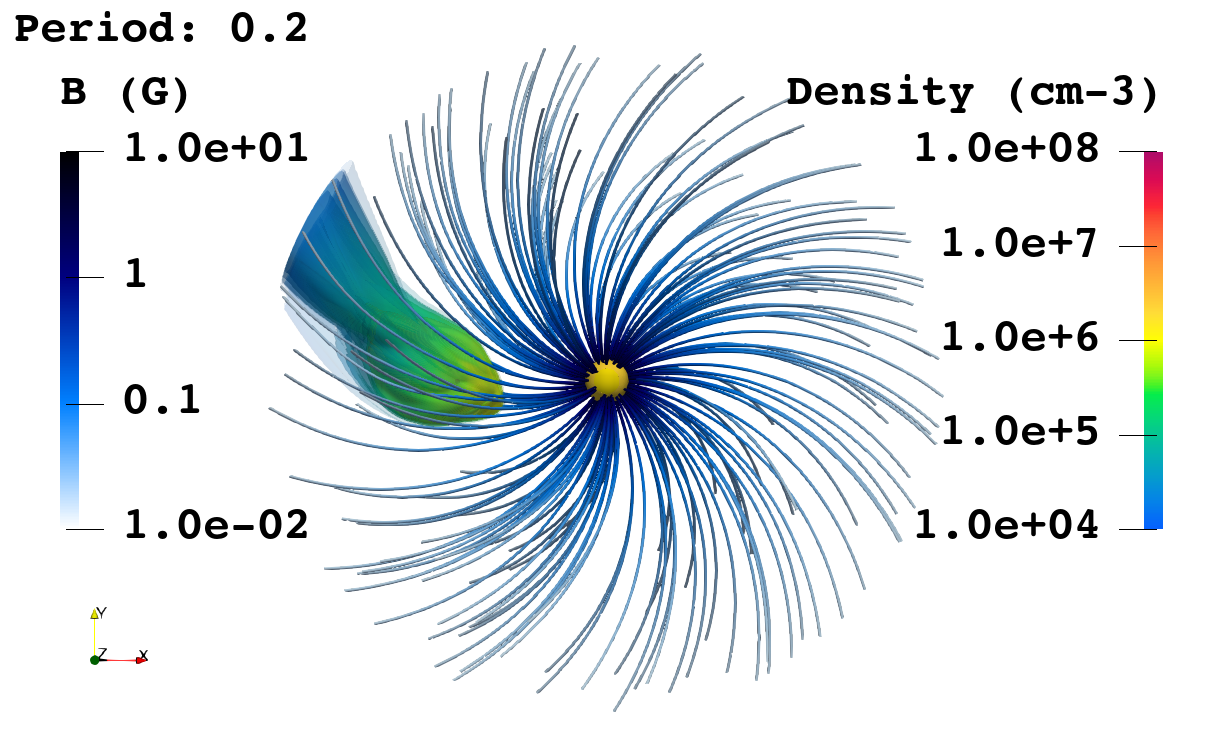}
    }
	\subfloat[ \label{fig:RefT20}]{
        \includegraphics[width=0.328\hsize]{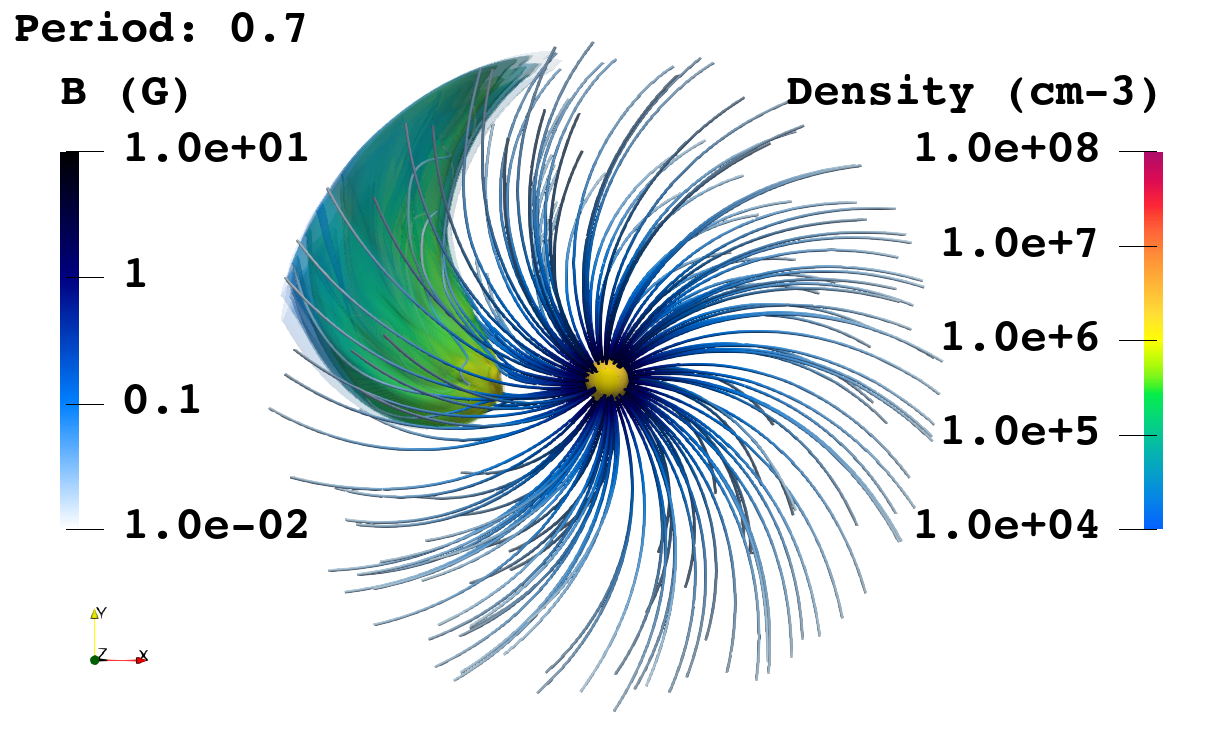}
    }
    \subfloat[\label{fig:RefT40}]{
        \includegraphics[width=0.328\hsize]{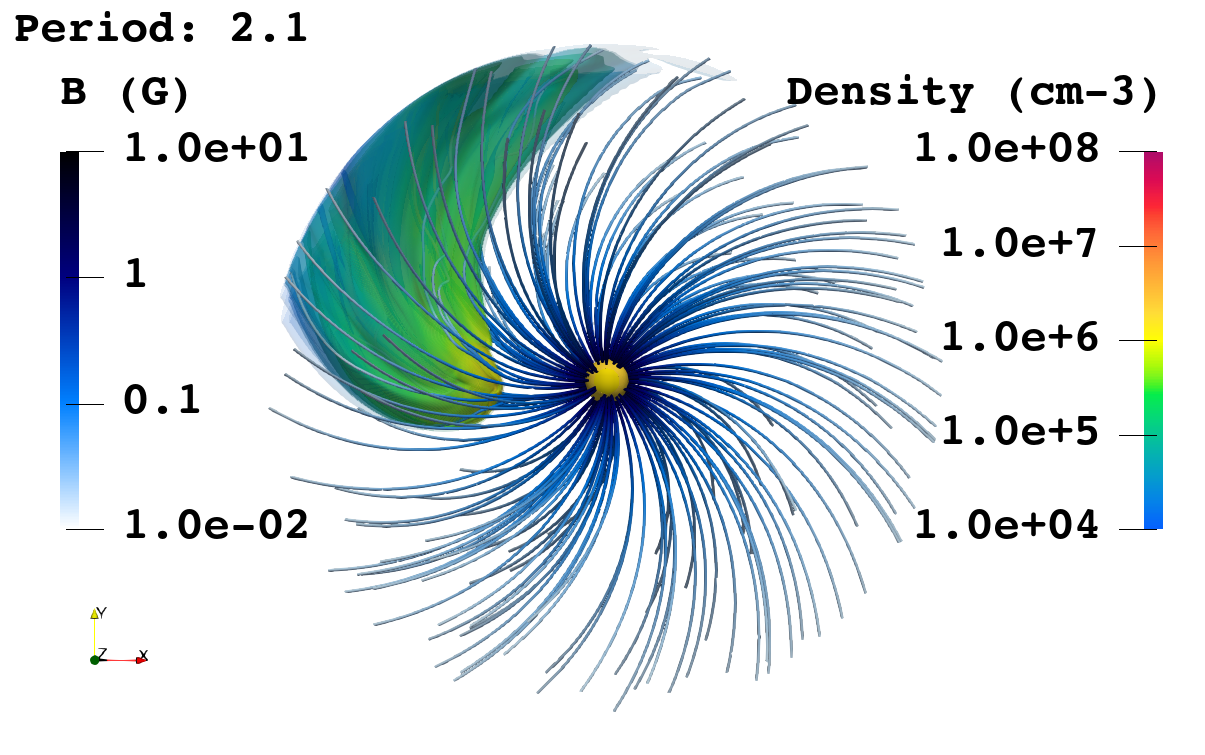}
    }

    \subfloat[\label{fig:RefT90} ]{
        \includegraphics[width=0.328\hsize]{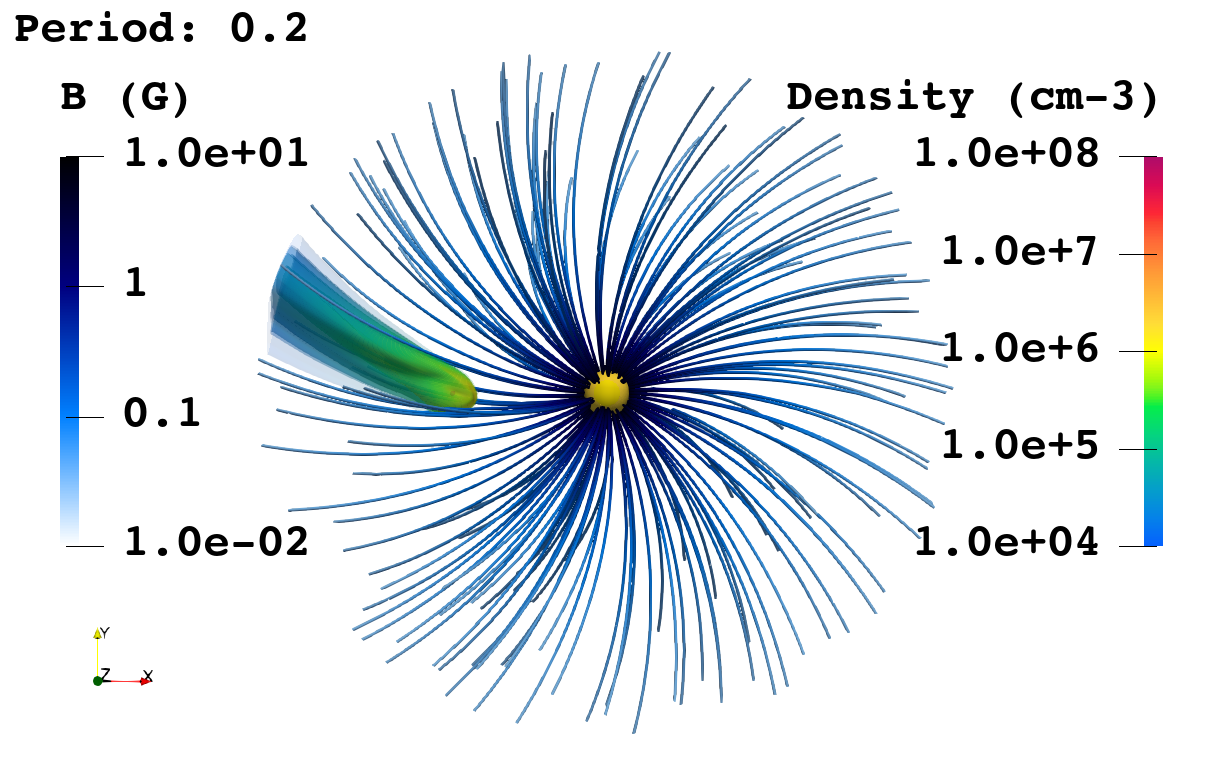}
    }
	\subfloat[ \label{fig:RefT120}]{
        \includegraphics[width=0.328\hsize]{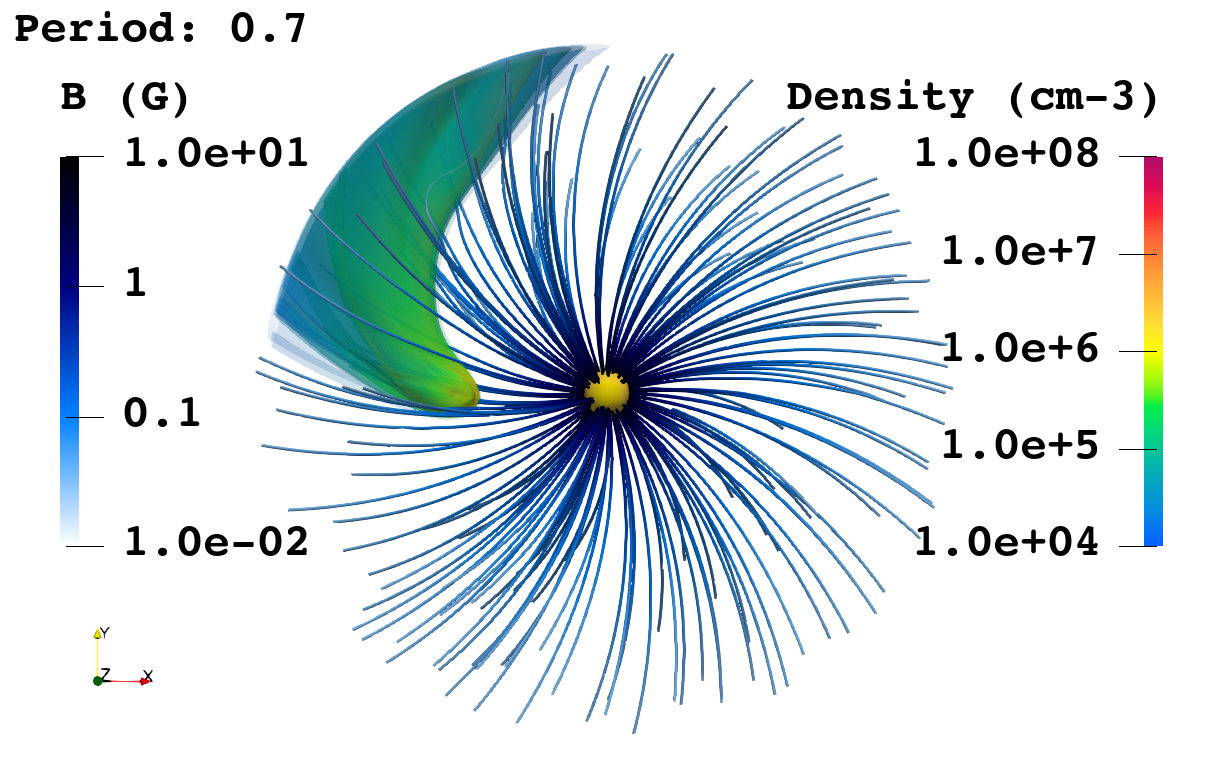}
    }
    \subfloat[\label{fig:RefT160}]{
        \includegraphics[width=0.328\hsize]{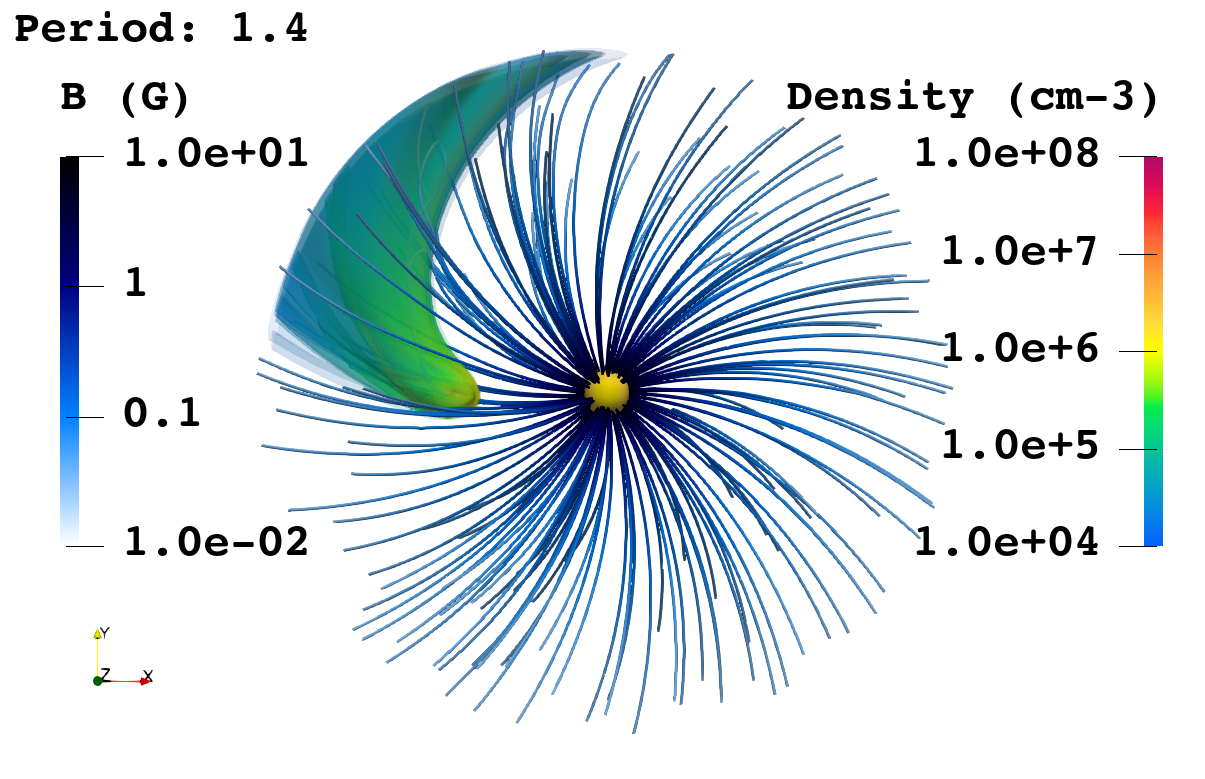}
    }
\caption{Evolution of density for the Bs5-Bp5 case (top three panels) and for Bs10-Bp1 (bottom three panels). The images show the density of the planetary wind in a log color scale. The yellow sphere at the centre represents the central star.}
\label{Fig:highBcases}
\end{figure*}

The dynamics exhibited by models with more intense magnetic fields compared to the reference case reveal simpler and distinctive features. These two cases provide compelling evidence for the fundamental role of the magnetic field in shaping the dynamics of the planetary wind. Fig. \ref{Fig:highBcases} illustrates the dynamics observed in the cases of Bs5Bp5 and Bs10Bp1.
Unlike the reference case, both Bs5Bp5 and Bs10Bp1 exhibit the absence of an accretion column. Instead, the planetary wind expands, assuming a droplet-like structure that surrounds the planet. 
In fact, the expansion of the planetary wind is impeded in these cases by the ram pressure of the stellar wind, presents also in all cases, and the magnetic pressure. 
Additionally, the intense magnetic field suppresses the Rayleigh-Taylor instabilities, which are responsible for the formation of the accretion column in the reference cases. This suppression prevents the material in front of the planet from accreting onto the star and instead pushes it behind, forming a cometary tail that extends from the planet. The material in this tail eventually escapes from the system.
Due to the interaction with the stellar wind, also in these cases, a shock forms due to the impact between the planetary and stellar wind. In particular, we observe a bow shock in front of the planet as a result of the interaction between the planetary wind and the stellar wind. This kind of dynamic has been adopted as an explanation for Ly$\alpha$ observation \citep[e.g.][]{Vidotto2011a,Vidotto2011b}

\begin{figure*}
\centering
    \subfloat[\label{fig:HDrho10} ]{
        \includegraphics[width=0.45\hsize]{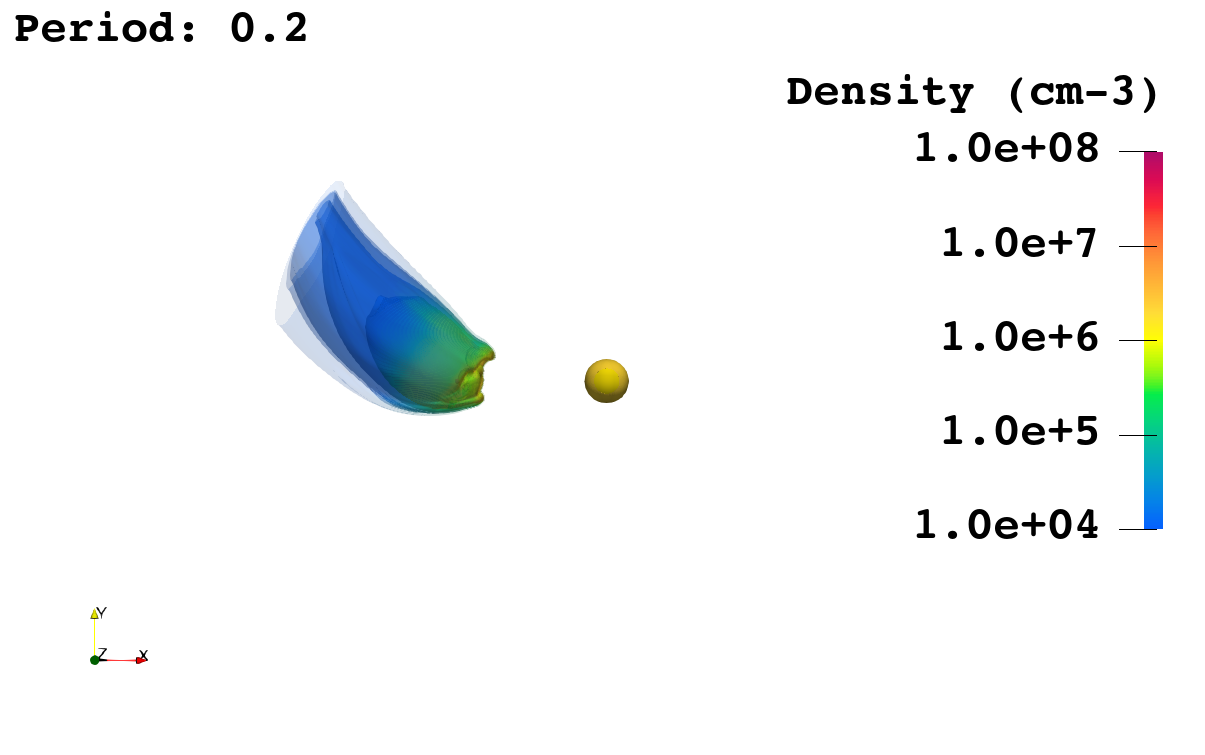}
    }
	\subfloat[ \label{fig:HDrho20}]{
        \includegraphics[width=0.45\hsize]{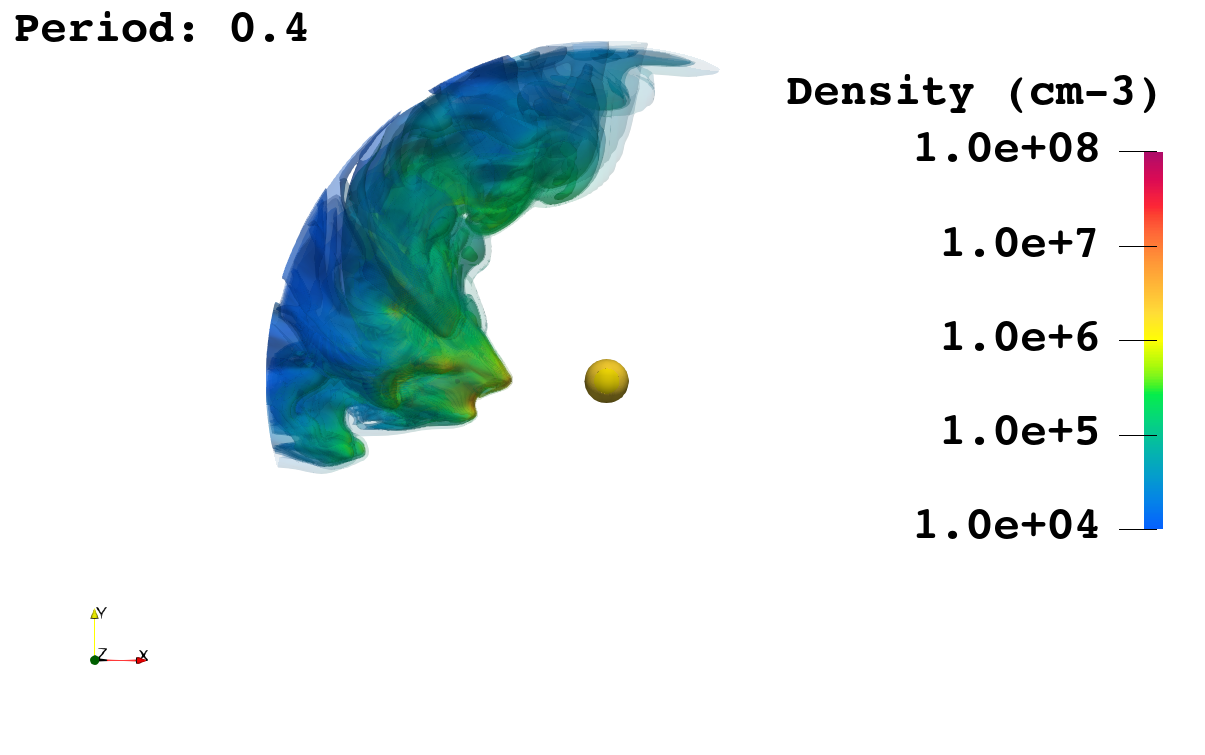}
    }
    
    \subfloat[\label{fig:HDrho40}]{
        \includegraphics[width=0.45\hsize]{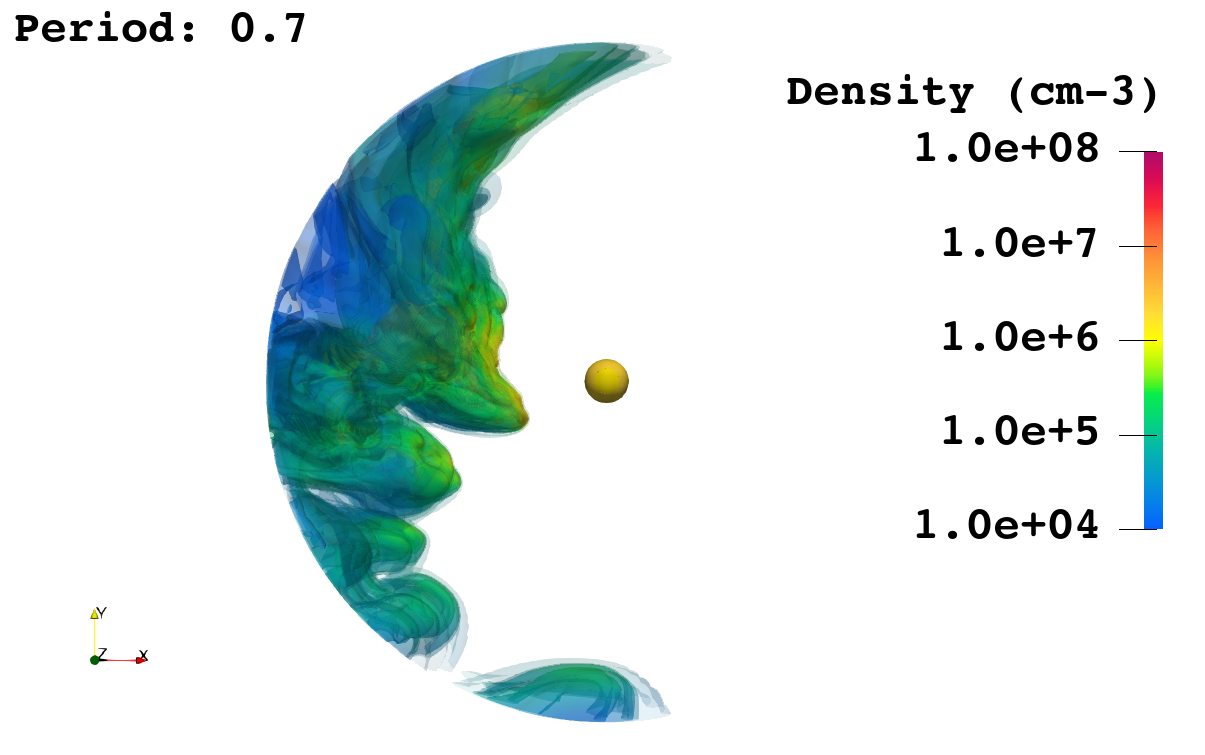}
        }    
	\subfloat[\label{fig:HDrho90} ]{
        \includegraphics[width=0.45\hsize]{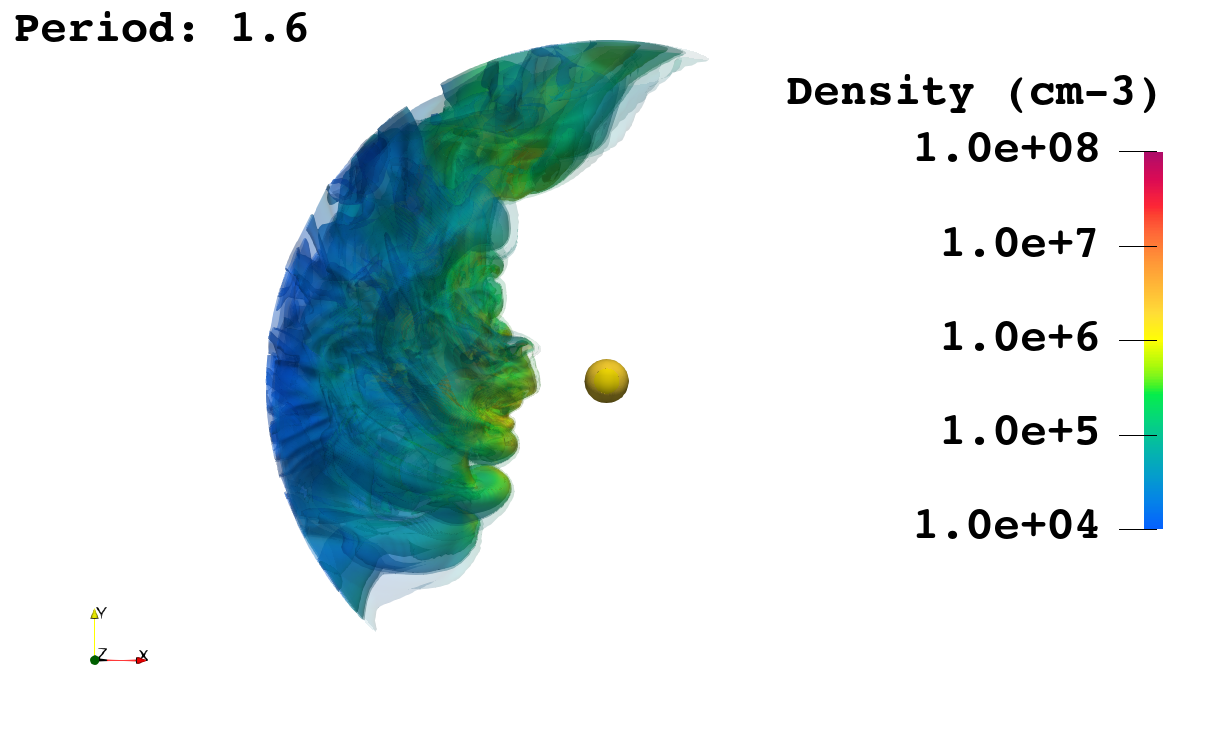}
    }
    
	\subfloat[ \label{fig:HDrho120}]{
        \includegraphics[width=0.45\hsize]{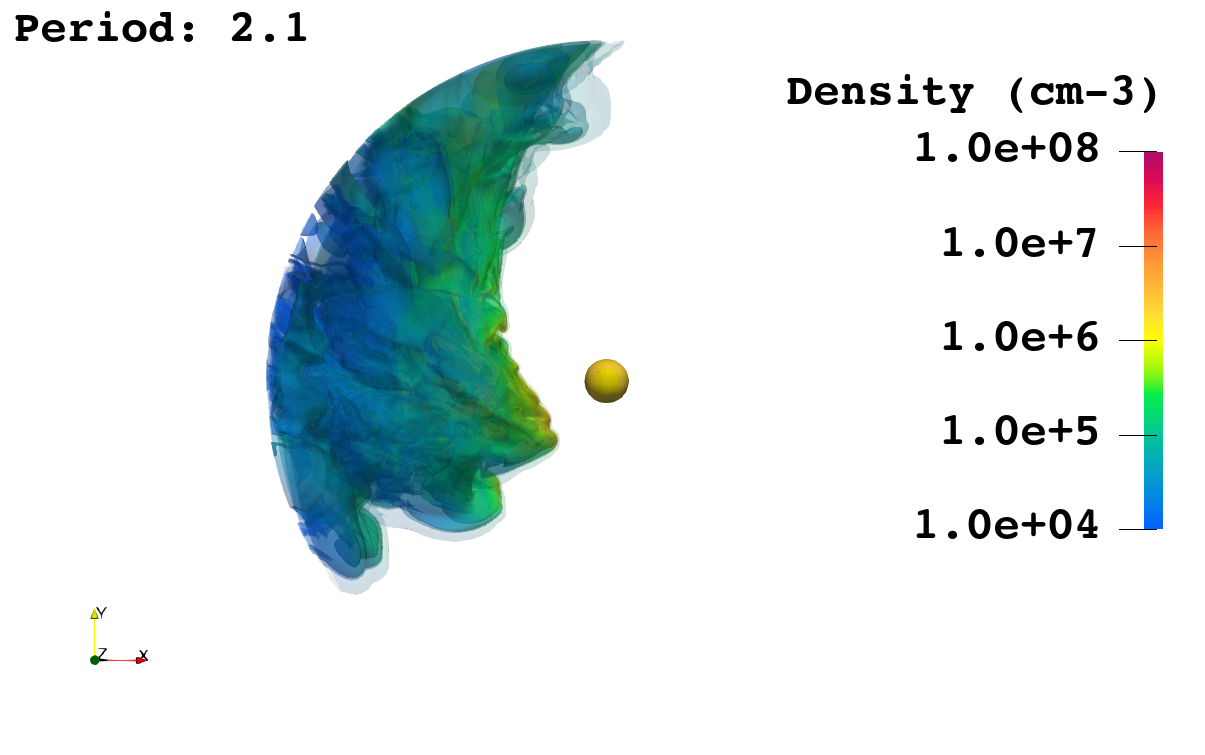}
    }
    \subfloat[\label{fig:HDrho160}]{
        \includegraphics[width=0.45\hsize]{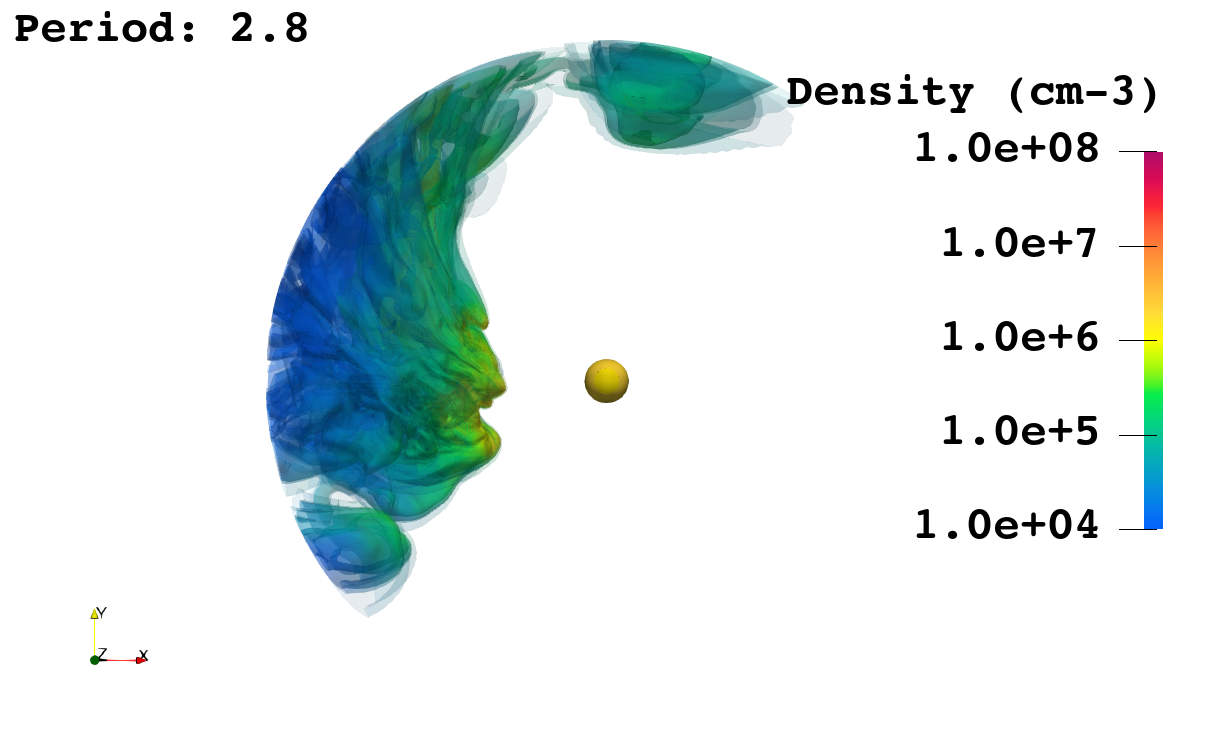}
    }

\caption{Evolution of density for the HD case. The images show the density of the planetary wind in a log colour scale. The yellow sphere at the centre represents the central star.}
\label{Fig:HDevo}
\end{figure*}

To uniquely identify the role of the magnetic field in determing the dynamics of the planetary wind, we also performed a purely hydrodynamic simulation (HD in Tab. \ref{tab:sims}), which is the case for B$_S$ = $B_p$ = 0. The evolution of the HD case is shown in Fig. \ref{Fig:HDevo}. 
In this case, the system's evolution closely resembles the reference case, but with one notable difference: no accretion is observed on the stellar surface. In fact, as the reference case, the planetary wind initially expands; during the expansions due to the interaction with the stellar wind Rayleigh-Taylor instabilities develop. The number of these instabilities is larger compared to the reference case and as a result, the planet is surrounded by a more extended cloud of planetary wind with respect to the reference case (see for comparison Fig. \ref{fig:HDrho120} and Fig. \ref{fig:RefD120}). 
In this particular case, the absence of accretion can be attributed to the material being dispersed more efficiently, primarily driven by a higher occurrence of Rayleigh-Taylor instabilities compared to the reference case. These increased instabilities contribute to the effective scattering and distribution of the material, preventing significant accretion on the stellar surface.

\subsection{Observability}

In this section we present the results of post processing of the data cubes of the simulations and the observables that can trace back to SPI effects in X-rays.
One of the goals of this work is to verify if the interpretation of \cite{Pillitteri2015} about the X-ray flares observed in HD189733 are the results of SPI, and in particular of the planetary wind hitting the stellar surface forming high-temperature X-ray emitting shocks. 
\subsubsection{Hotspots}
The simulations show that the only case in which we observe accretion and thus hotspots onto the stellar surface is the reference case with $B_p = 1G$ and $B_s = 5G$.
\begin{figure}[!h]
\centering
    \subfloat[\label{fig:HS90} ]{
        \includegraphics[width=0.9\hsize]{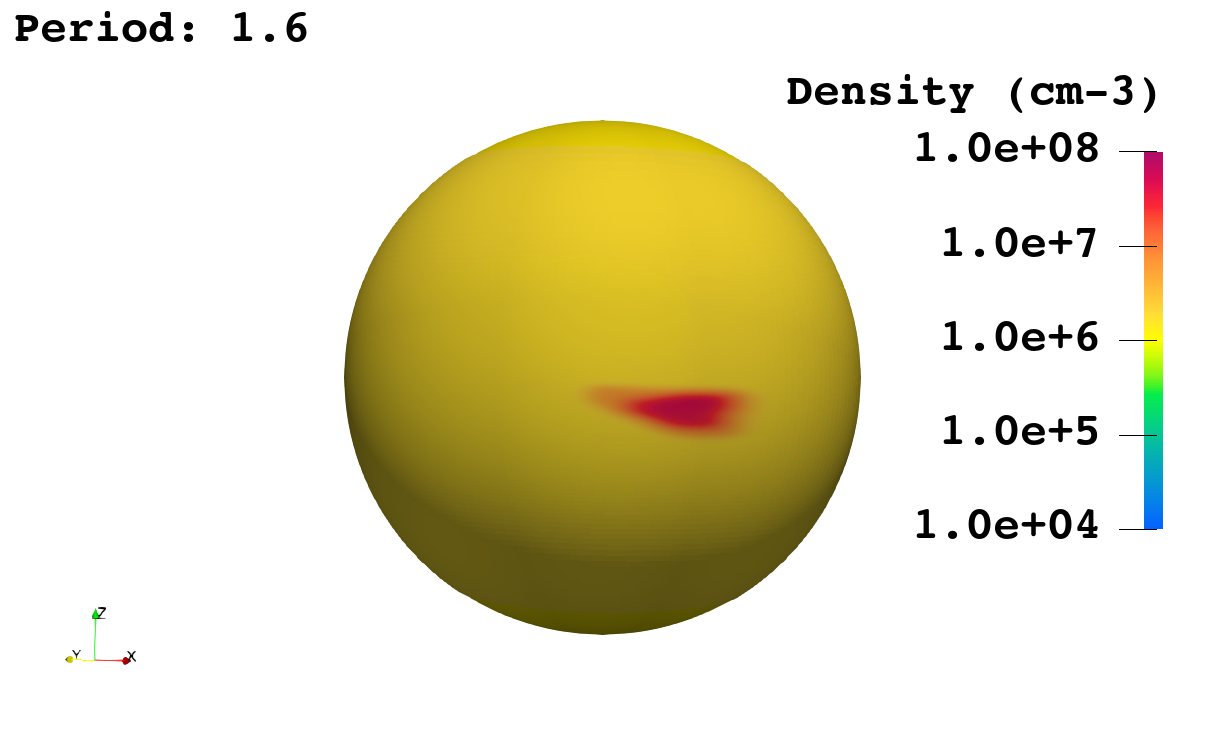}
    }
    
	\subfloat[ \label{fig:HS120}]{
        \includegraphics[width=0.9\hsize]{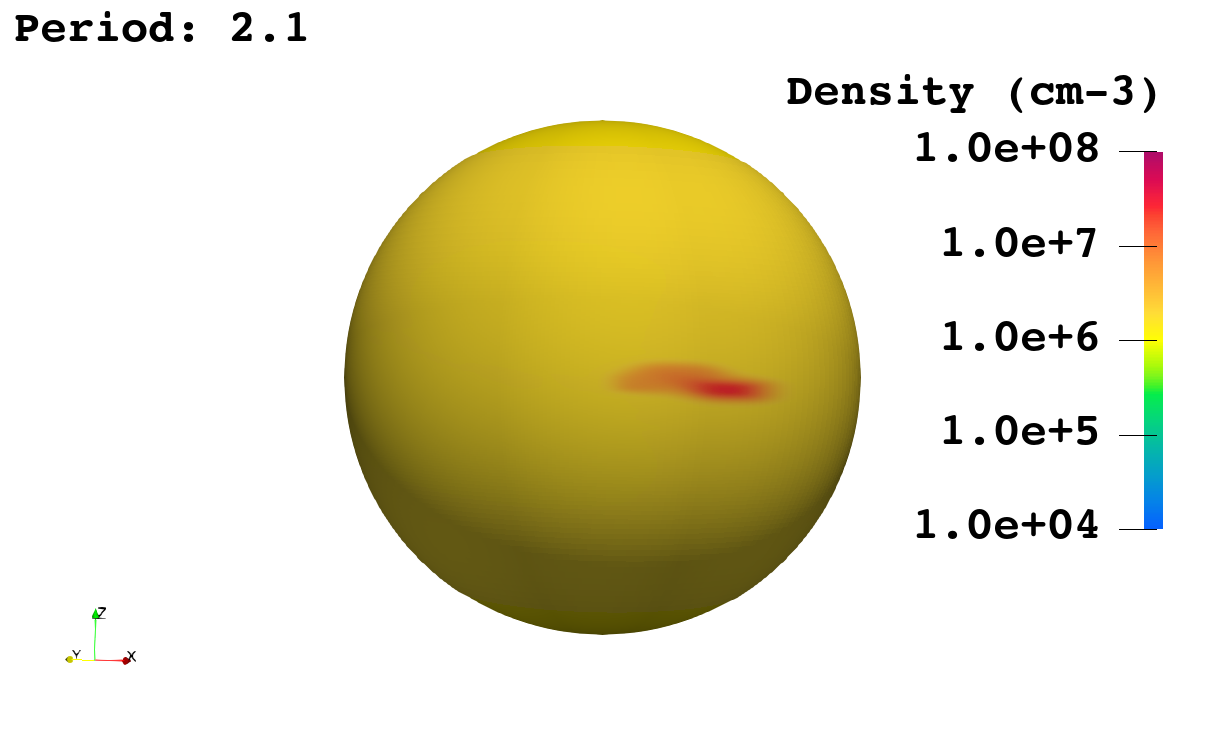}
    }
    
    \subfloat[\label{fig:HS160}]{
        \includegraphics[width=0.9\hsize]{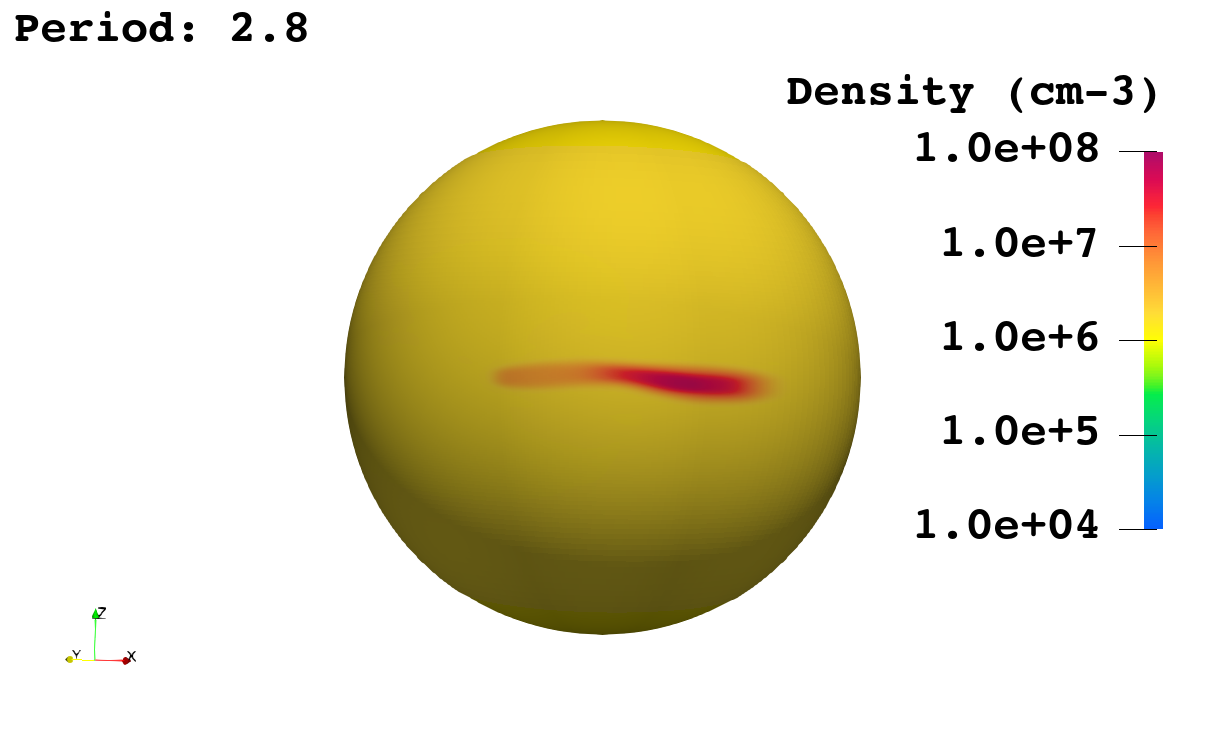}
        }
    \caption{Map of density in log scale at the stellar surface. The spot is located at about $45^\circ-60^\circ$ ahead of the planet. \label{fig:HSEvo}}
\end{figure}
Fig. \ref{fig:HSEvo} shows the regions where we expect to observe the impact of the material on the stellar surface and the subsequent generation of hotspots. As expected, the hotspot mirrors the dynamics of the planetary wind. In fact, once the accretion column is formed, the hotspot remains stable throughout the entire simulation. It is worth noting that this model does not describe the shocks due to the impact of the accreting gas on the stellar surface. In fact, the model does not include a proper description of the stellar atmosphere which is assumed to be an external boundary (See Sect. \ref{sect:model}).
Nonetheless, we can discuss the expected dynamics and the location of the hotspots and their observability. In particular, the hotspots appear to be close to the stellar equator, as a result of the Parker spiral magnetic field configuration assumed. 
%A different magnetic field (e.g.: an octupole component) would result in a higher latitude for the accretion column. However, the exploration of different magnetic field topologies is not the goal of this work.

The amount of X-ray emission emitted by an accretion shock for an optically thin plasma is proportional to the post-shock plasma density squared. Under the assumption of strong shock \citep{Zeldovich2012}, $\rho_{ps} = 4\rho_{accr}$. For these reasons, in order to study the observability we estimate the accretion rate of material into the stellar surface and the maximum density on the accretion column.
\begin{figure}[!htbp]
    \centering
    \includegraphics[width=\hsize]{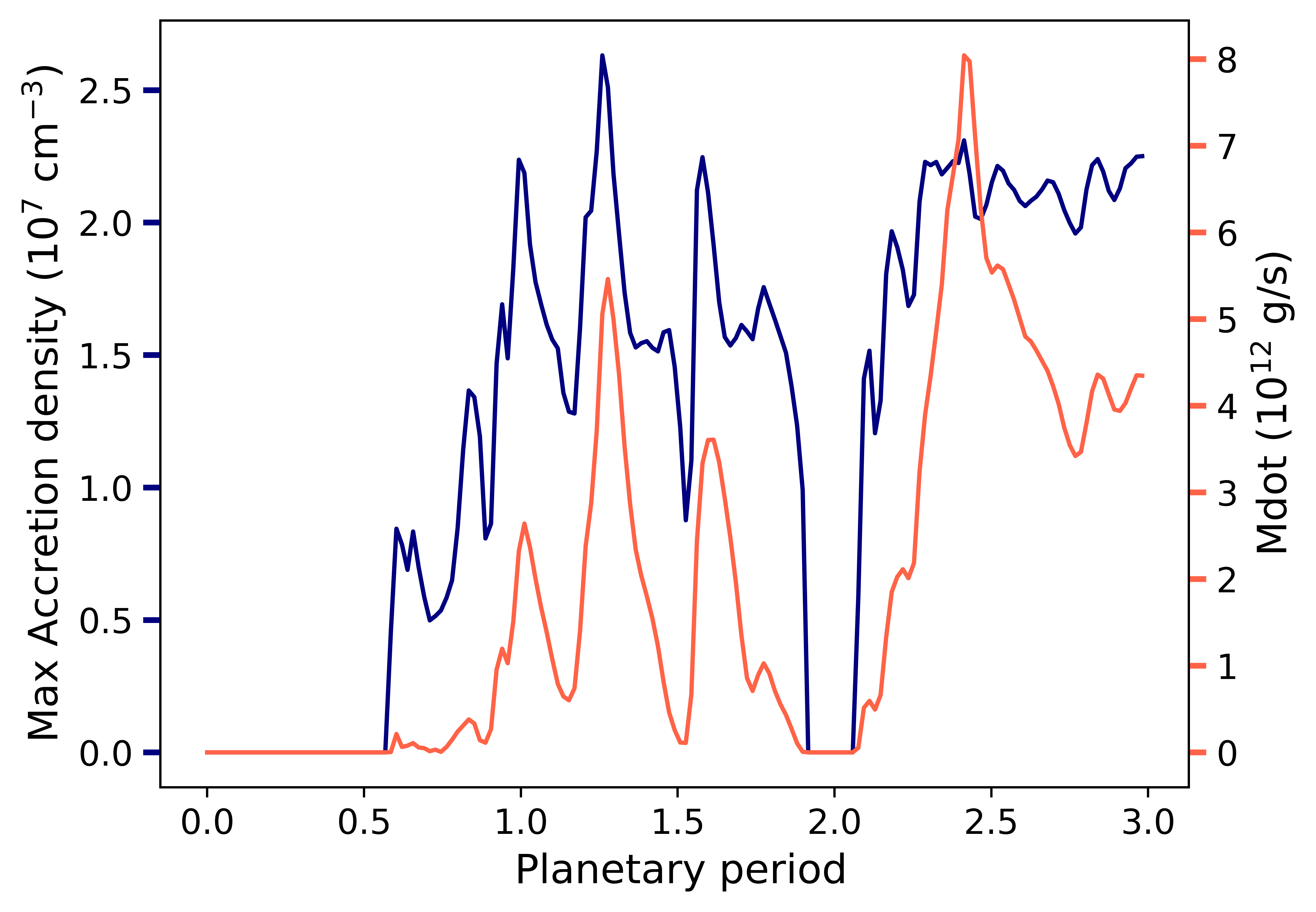}
    \caption{Maximum value of density on the accretion column (in blue) and accretion rate (in red) vs planetary period.}
    \label{fig:AccRate}
\end{figure}
As Fig.\ref{fig:AccRate} shows, after the initial transient phase that it takes about 1 orbital period to establish an accreting column out of the planetary outflow, 
the accretion rate increases to $10^{12}$g s$^{-1}$ with a maximum value of $8\times 10^{12}$g s$^{-1}$, this value is dramatically lower than the accretion rates measured in CTTSs that are typically of the order of $10^{17}$ g s$^{-1}$ \citep{Hartmann2016}. The lower accretion rate compared to the CTTSs is expected, in fact in this case the source of accretion is the planet which is evidently a smaller reservoir of mass compared to the stellar disk.
For a complete analysis of the observability of the accretion events we also synthesize the maximum density on the accretion column, which is shown in Fig. \ref{fig:AccRate}. The maximum density is about $10^7$cm$^{-3}$ with a maximum value at $2.5\times 10^7$ cm$^{-3}$. According to \citep{Zeldovich2012} this corresponds to a post-shock density of $10^8$cm$^{-3}$ which is one order of magnitude lower than typical coronal densities, which means that the emission produced would be indistinguishable from the typical coronal emission. 
It is important to stress that we are in the most favourable conditions to produce accretion. In fact, we are assume an extremely high evaporation rate from the planet (See Section \ref{sect:model}). For this reason we can conclude that the accretion shock due to SPI can be excluded as the source of X-ray flare in phase with the planet observed by \cite{Pillitteri2015}. 

\subsubsection{X-ray emission from diffused wind}
From the results of the models, we synthesize the X-ray emission arising from the planetary wind (see Sect. \ref{sect:em}). In the previous section we demonstrate that flare-like emission observed by \cite{Pillitteri2015} cannot be originated from the impact of the planetary material onto the stellar surface. Here we discuss if the flare-like emission originated from the hot wind surrounding the planet.

\begin{figure}
    \centering
    \includegraphics[width=\hsize]{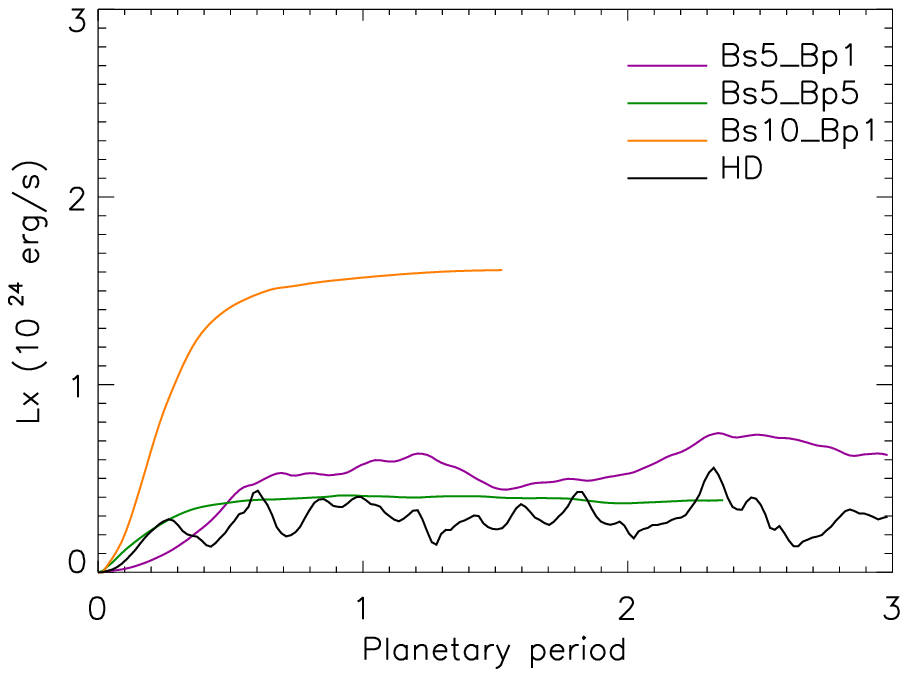}
    \caption{X-ray luminosity vs planetary period for different cases. In purple is the Bs5\_Bp1 case, in green the Bs5\_Bp5 case, in yellow the Bs10\_Bp1 case and in black the HD case.}
    \label{fig:Xlum}
\end{figure}

Figure \ref{fig:Xlum} illustrates the X-ray light curves obtained from different simulations. In each case, there is a noticeable increase in emission during the transient phase, which spans approximately half of the planetary period, leading the system into a stable dynamical regime. Subsequently, the emission remains stable, but the HD and Bs5\_Bp1 cases exhibit more significant variability compared to the other two cases. This variance is attributed to the weaker magnetic field, which does not effectively suppresses instabilities. In fact, the cases Bs5\_Bp5 and Bs10\_Bp1 that show stronger magnetic fields present no further variability once they reach the stationary regime.

In all the cases analysed, the total emission is on average between 4 orders of magnitude lower than the average X-ray luminosity for HD189733b \citep[$\approx 10^{28}$erg s$^{-1}$ e.g.][]{Pillitteri2022}. However, it is worth noticing that the case with the strongest magnetic field is the one that shows the highest value of X-ray emission. This is due to the fact that the magnetic confinement of the plasma works at its highest efficiency in the models we explored, and produces the highest density region enshrouding the planet.  

Even in this case, we can conclude that this is not the origin of the emission observed by \cite{Pillitteri2015}.

\section{Conclusions}\label{conclusions}

In this work, we analysed whether the X-ray emission which appears in phase with the planetary period observed in HD189733b by \cite{Pillitteri2015} is a result of star-planet interaction occurring on HD189733b.
In particular, we aimed at verifying whether the planetary wind may be responsible for the flare-like emission observed. For this reason, we developed a 3D MHD model that describes the system HD189733A and its hot Jupiter with extremely high planetary wind; we explored different magnetic field intensities.

The results of this work are the following: 
\begin{itemize}
    \item The planetary wind expands and interacts with the stellar wind and stellar magnetic field. During the expansion the impact with the stellar wind generates Rayleigh-Taylor instabilities that, if strong enough, push the planetary wind out of equilibrium. The planetary wind is then caught by the stellar gravity and accretes onto the stellar surface forming an accretion-column-like structure that links star and planet.
    \item The accretion onto the star of the planetary wind is not common. In 4 cases explored, the material accretes onto the stellar surface only in 1 case. In fact, in order to fall into the star we need a good combination of magnetic field intensities. A lower magnetic field intensity does not suppress efficiently the Rayleigh-Taylor instabilities leading to a more diffuse cloud of planetary wind. On the contrary, a higher magnetic field intensity suppresses completely the Rayleigh-Taylor instabilities which trigger the accretion column.
    \item Even in the case that shows accretion the impact region cannot be responsible for the X-ray variability observed by \cite{Pillitteri2015}. In fact, the density of the material that hits the stellar surface is of the order of $10^7$ cm$^{-3}$ which is too low to generate an emission distinguishable from the typical coronal emission that, for the solar case if of the order of $10^8$ cm$^{-3}$.
    \item The interaction of the planetary wind with the stellar wind generates bow shocks that produce a high-temperature region between the star and the planet. However, the X-ray emission generated from this region is of the order of $10^{24}$ erg/s which is negligible compared to the average luminosity observed in HD189733A which is about $10^{28}$ erg/s \citep[e.g.][]{Pillitteri2022}.

\end{itemize}

We can conclude that this kind of SPI can not be responsible for the X-ray emission bursts observed in HD189733A.
Other work claims that there is no statistical evidence for a bright hot spot synchronized to the planetary period and that the observed variability on HD 189733A is compatible with the normal evolution of active regions on star \citep{Route2019}. 

However, SPI that involves accretion onto the stellar surface and interaction between planetary and stellar wind it is not the only scenario that can produce the X-ray flares in phase with the planetary period. It is important to notice that the magnetic configuration in the tail of the planetary wind is highly perturbed. In principle, the interaction of the planetary wind with the stellar magnetic field can generate clumpy regions, as observed for the solar case \citep{Petralia2016}. The perturbed magnetic field configuration may also generate recombination effects that may heat up the clumpy regions and produces observable X-ray emission. In order to investigate this effect and its observability, a more spatially resolved model is required with the addition of a resistive term that takes into account the heating of the plasma due to magnetic reconnection.

\begin{acknowledgements}
We acknowledge the financial contribution from the agreement
ASI-INAF No. 2018-16-HH.0 (THE StellaR PAth project). We acknowledge support from ASI-INAF agreement 2021-5-HH.0 Partecipazione alla fase B2/C della missione ARIEL (Atmospheric Remote-Sensing Infrared Exoplanet Large survey).
PLUTO was developed at the Turin Astronomical Observatory and the Department of Physics of Turin University. We acknowledge the HPC facility (SCAN) of the INAF – Osservatorio Astronomico di Palermo for the availability of high-performance computing resources and support. We acknowledge PRACE for awarding access to the Fenix Infrastructure resources at CINECA, which are partially funded from the European Union’s Horizon 2020 research and innovation programme through the ICEI project under the grant agreement No. 800858.
\end{acknowledgements}

\bibliographystyle{aa} 
\bibliography{biblio.bib}

\section{Appendix: Resolution convergence test}\label{Appendix}

\begin{figure}
     \includegraphics[width=\hsize]{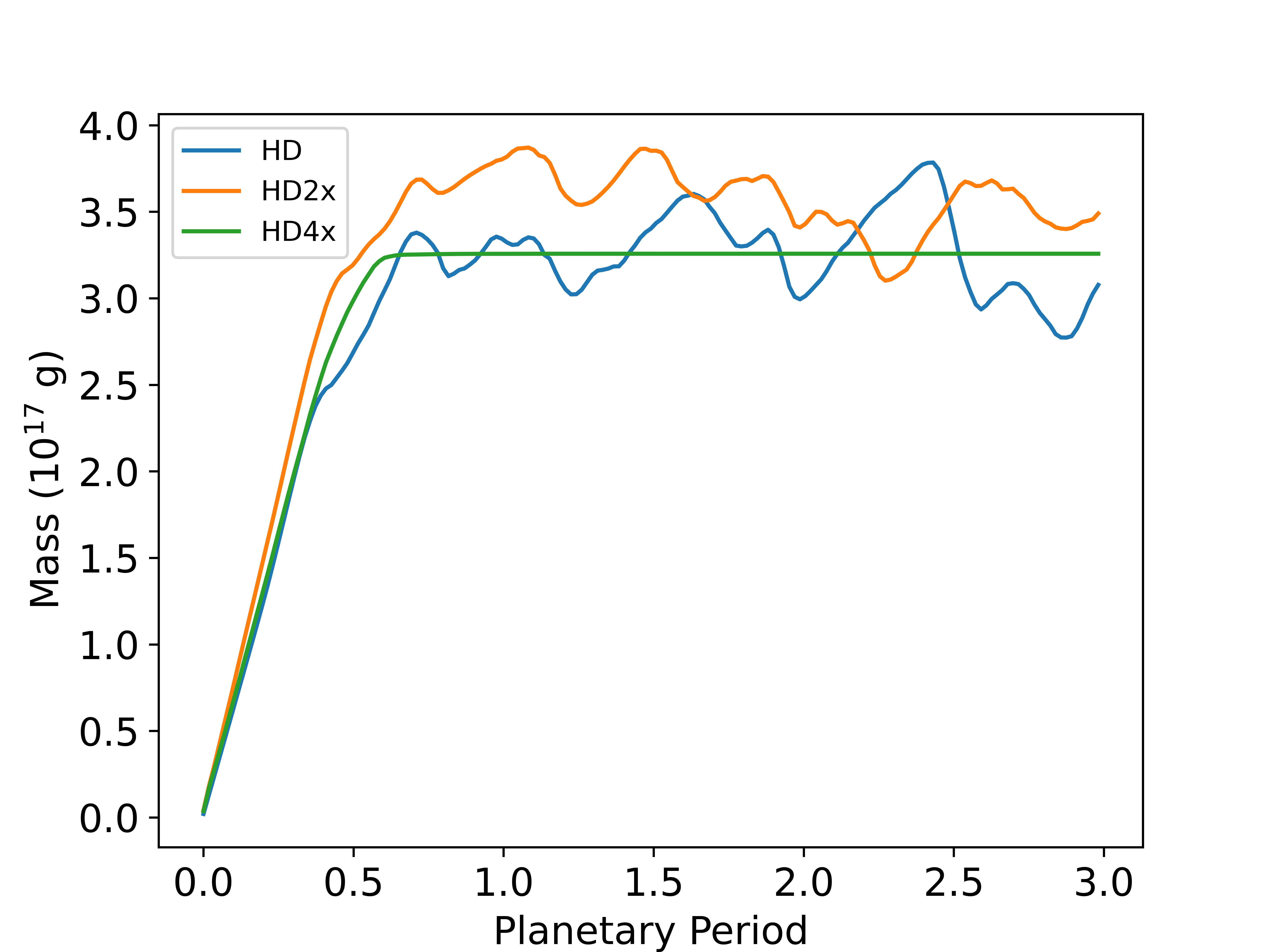} 
\caption{Mass of planetary outflow (at least 10\% of planetary origin) as a function of time. The green line is the HD4x case, the yellow line is the HD2x case and the blue line is the HD case.}
\label{Fig:ConvTest}
\end{figure}

The spatial resolution adopted reflects a compromise between the need of high resolution and the computational cost of each simulation. 
To achieve this goal, we performed a convergence test on the HD case, considering two additional simulations with numerical meshes $4\times$ (HD4x) and $2\times$ (HD2x) coarser than the one adopted in the paper, respectively. 
Indeed, this case is characterized by RT instability, which results in the formation of small-scale structures that may be influenced by the numerical grid and, consequently, by the spatial resolution. The convergence test was performed through the tracer defined for the planetary wind to identify zones whose content is made up of original planetary wind material by more than 10\%. Figure~\ref{Fig:ConvTest} shows the total mass of these zones in the computational domain during the evolution in the three cases analyzed.

We found that the results obtained with different spatial resolutions exhibit the same general trend, with differences not exceeding 15\%. After the initial transient phase of about half orbital period, during which the mass linearly increases with time as the planetary wind expands through the spatial domain, a stationary condition is achieved around a value of $M_{\rm pw} \approx 3 \times 10^{17}$~g. This suggests that an equivalent amount of mass enters the domain from the planet's surface and exits from the external boundary of the domain. On the other hand, we note a striking difference between case HD4x and the other two cases: after the initial transient, the total mass $M_{\rm pw}$ remains constant in HD4x, whereas it exhibits chaotic and similar variability in the other two cases. This is attributed to the onset of RT instability, leading to substantial perturbations in the planetary wind. Remarkably, these perturbations appear very similar in the two cases HD2x and HD, indicating that our simulations successfully capture the fundamental properties of the dynamics. In light of this, we decided not to further increase the spatial resolution and opted for the HD one.

\end{document}